\documentclass[useAMS,usenatbib]{mn2e}
\usepackage{amsmath}
\usepackage{epsfig}
\newcommand{\apj}{ApJ}

\newcommand{\pasp}{PASP}
\title[Molecular cloud regulated star formation in galaxies]{Molecular cloud regulated star formation in galaxies}

\author[C. M. Booth, Tom Theuns and Takashi Okamoto]{C. M. Booth$^{1}$\thanks{E-mail: c.m.booth@durham.ac.uk (CMB)}, Tom Theuns$^{1,2}$ and Takashi Okamoto$^{3}$\\
$^{1}$Institute for Computational Cosmology, University of Durham, South Road, Durham DH1 3LE\\
$^{2}$University of Antwerp, Campus Groenenborger, Groenenborgerlaan 171, B-2020 Antwerpen, Belgium\\
$^{3}$Division of Theoretical Astronomy, National Astronomical Observatory of Japan, 2-21-1 Osawa, Mitaka, Tokyo 181-8588, Japan}

\begin{document}

\pagerange{\pageref{firstpage}--\pageref{lastpage}} \pubyear{2006}

\maketitle

\label{firstpage}

\begin{abstract}
We describe a numerical implementation of star formation in disk
galaxies, in which the conversion of cooling gas to stars in the
multiphase interstellar medium is governed by the rate at which
molecular clouds are formed and destroyed. In the model, clouds form
from thermally unstable ambient gas and get destroyed by feedback
from massive stars and thermal conduction. Feedback in the ambient
phase cycles gas into a hot galactic fountain or wind. We model the
ambient gas hydrodynamically using smoothed particle hydrodynamics
(SPH). However, we cannot resolve the Jeans mass in the cold and dense
molecular gas and, therefore, represent the cloud phase with ballistic
particles that coagulate when colliding. We show that this naturally
produces a multiphase medium with cold clouds, a warm disk, hot
supernova bubbles and a hot, tenuous halo.  Our implementation of
this model is based on the Gadget N-Body code. We illustrate the model
by evolving an isolated Milky Way-like galaxy and study the
properties of a disk formed in a rotating spherical collapse. Many
observed properties of disk galaxies are reproduced well, including
the molecular cloud mass spectrum, the molecular fraction as a
function of radius, the Schmidt law, the stellar density profile and
the appearance of a galactic fountain.
\end{abstract}


\section{Introduction}
\label{sec:intro}
Galaxies form when gas cools radiatively inside a dark matter (DM)
halo. The haloes in turn form from the non-linear collapse of
initially small density perturbations that were imprinted by quantum
fluctuations before inflation. When the virial temperature of the halo
is high enough the gas can cool radiatively (\cite{rees77,whit78}) and
may become self-gravitating. Some fraction of the cooling gas can form
stars, which then affect the baryon distribution and star formation
rate through feedback; for example from supernova (SN) explosions
(e.g. \cite{deke86}).

Simulating the growth of dark matter haloes from initially small
cosmological density perturbations has become routine to the extent
that even the complex non-linear stage can be predicted with relative
confidence (e.g. \cite{spri05a}). However, following the behaviour of
the baryons until stars form is much more demanding for two main
reasons. Firstly, the physical processes that lead to star formation
are still relatively poorly understood, even in the Milky Way
(MW). Stars are thought to form in molecular clouds in a complex
interstellar medium (ISM) in which magnetic fields (\cite{safi97}),
cosmic rays, turbulence (\cite{krum05}), relativistic jets
(\cite{klam04}), molecules, dust, and radiative transfer may all play
some role. Secondly, the physical scales on which star formation takes
place is vastly different from those of cosmological interest.
Therefore, simulations on a sufficiently large scale to sample
cosmological structures cannot presently also resolve physics on the
scales relevant for star formation. For these two reasons simulations
invariably include \lq sub-grid\rq\, physics that model the complexity
of star formation in the interstellar medium using simple rules.

In the first generation of hydrodynamic models for galaxy formation
(\cite{nava93,stei94,cen92,katz96}), the gaseous component of galaxies
is modelled as a single fluid.  Because the spatial and mass
resolution in these simulations is insufficient to resolve star
formation they rely upon a simple star formation prescription. In its
simplest form this consists of devising criteria by which simulated
gas can be flagged as eligible to form stars and then converting the
gas to stars by hand (see e.g. \cite{kay02} for a comparison of such
criteria). Feedback from core collapse (type II) SNe, modelled as an
extra source of thermal or kinetic energy, was found to have little
effect in the early models. This is because the gas in the
surroundings of star formation sites is at high density and so is very
efficient at just radiating away the added energy. As a consequence,
too much gas cooled in dense knots, producing galaxies much more
concentrated than those observed (\cite{nava91,weil98}).

The ISM in observed galaxies is much more complex than the single
phase medium present in these early simulations. Although observed
galaxies may have comparable mean gas density to the simulated ones,
the real ISM consists of very dense cold clouds with small volume
filling factor, embedded in a much more tenuous low-density, hot
medium with a \lq warm\rq\ phase at the interface. SN explosions in
the tenuous medium have a much greater impact on the galaxy because
this medium cools much less efficiently than gas at the mean density.
Throughout this work we will use the label \lq cold\rq~ to describe
the molecular clouds in the ISM; and the labels \lq warm\rq~ and \lq
hot\rq~ to describe the properties of the ambient gas phase at
temperatures of approximately $10^4$K and $10^6$K respectively.

In response to these problems with the simplest star formation and
feedback criteria several authors have introduced \lq multiphase\rq\
models for star formation in which the ISM is treated as a number of
distinct phases.  These schemes take various forms including
modification of the simulation algorithm
(\cite{ritc01,crof00,scan06}), treating the multiphase medium
implicitly by formulating differential equations that model the
interactions between the phases (\cite{yepe97,spri03,okam05}),
explicitly decoupling supernova heated gas from its surroundings
(\cite{stin06}), or by decoupling the cold molecular phase from the
hot phase by one of a variety of methods, including \lq sticky
particles\rq\ (\cite{seme02,harf06}) or removing \lq cold\rq\
particles from the SPH calculation
(\cite{hult99,pear99,pear01,marr03}).  The decoupling of the hot and
cold ISM phases allows thermal heating from SN feedback to become more
efficient (due to the much lower density of the hot phase) and also
allows one to follow the properties of the cold molecular phase of the
ISM.

This paper describes an attempt to mimic the multiphase medium in a
star forming galaxy. Since stars are known to form in molecular
clouds our method focuses on simple rules for cloud formation. The
star formation recipe then simply converts the most massive of these
clouds, \lq Giant Molecular Clouds\rq~ (GMCs), into stars with an
imposed efficiency taken from MW observations. Once a GMC forms stars
stellar feedback destroys the cloud. We do not attempt to model the
clouds themselves using hydrodynamics, because our current galaxy
formation simulations lack by some margin the dynamic range to resolve
the Jeans mass ($M_J$) in the cold, dense gas that makes up the
clouds.  We can demonstrate this lack of resolution by considering
theoretical models of the ISM (e.g. \cite{wolf95}), who calculated the
thermal equilibrium gas properties of a diffuse ISM.  \cite{wolf95}
found that a stable two phase medium was produced, with a transition
from hot material at densities $<10^{-0.5} {\rm cm}^{-3}$ to cold,
molecular material at densities $>10^{1.5}{\rm cm}^{-3}$. The Jeans
mass of the warm ($T=10^4 {\rm K}$) phase is approximately $1\times
10^8 M_{\odot}$, whereas the Jeans mass of the cold molecular gas
($T=100 {\rm K}$) is $\sim 2000 M_{\odot}$.  Typical cosmological
galaxy simulations currently have mass resolutions no better than
$10^6 M_{\odot}$ (\cite{okam05}), many orders of magnitude away from
being able to resolve the relevant scales for accurate tracking of the
cold molecular phase. Note that these simulations do however resolve
$M_{\rm J}$ in the warm phase. Given this limitation, we are forced to
instead introduce another particle type in our simulations, called a
\lq sticky particle\rq, to represent the clouds. These move as
ballistic particles through the ambient medium, yet when they
encounter another sticky particle interact inelastically based on a
set of collision rules that mimic the coagulation of clouds. Sticky
particles have been used before for similar reasons by e.g.
\cite{lin76,jenk94} and \cite{seme02}, and seem to have been
introduced originally by \cite{lars78} and \cite{levi81}. We show
below that our imposed collision rules produce cloud statistics that
are similar to those determined in nearby galaxies, for which they can
be measured.

We begin by giving a brief overview of the current theory of the ISM
and the formation of clouds, on which our recipes are based (section
\ref{sec:obssf}). We then introduce the physics we model with the
sticky particle prescription (section \ref{sec:model}), and constrain
all of the free parameters in the model (section
\ref{sec:parameter-est}).  We then present results from simple
isolated galaxy simulations (section \ref{sec:results}) and
investigate the effects of changing the physics in the sticky particle
model (section \ref{sec:away}).  Section \ref{sec:conclusions}
contains conclusions and details of future work.  The ISM is a
complicated system and the nomenclature used to describe it is
correspondingly complex.  For this reason appendix
\ref{app:list-of-symbols} contains a list of symbols and their
associated meanings. Readers not interested in numerical details can
read the summary of the model in section \ref{sec:model} and then
directly proceed to the analysis of the properties of the simulated
galaxies presented in section \ref{sec:results}.

\section{Star formation in disk galaxies}
\label{sec:obssf}
\subsection{The interstellar medium in disk galaxies}
According to the models of \cite{mcke77} (hereafter MO77; see also
\cite{efst00,mona04,krum05}) the ISM of the MW consists of at least
three separate and distinct gas phases: a hot tenuous medium at a
temperature of $\sim 10^6K$; cold, dense molecular clouds at a
temperature of $\le 100K$ and a warm medium that exists at the
boundaries between clouds and the hot medium with $T\sim 10^4K$. In
the MW, the hot medium has a filling factor of 70--80\% and the cold
clouds account for a few percent of the volume (MO77). Different
techniques are used to observe these different phases, with radio
observations probing roto-vibrational transitions of molecules (mainly
CO), the 21-cm line probing atomic hydrogen, and UV- and X-ray
observations probing the hot phase, see e.g. \cite{binn98} for an
overview and further references. The fact that different techniques
are used to observe the different gas phases might exaggerate the
degree to which these phases are really distinct.

Observations of star formation in the MW show that most stars form in
groups, either as gravitationally bound clusters or unbound
associations, in the most massive of the molecular clouds (Giant
Molecular Clouds, hereafter GMCs), with masses $\sim 10^6M_\odot$, and
sizes of order 30-50pc, see e.g. \cite{blit80} and \cite{lada03} for
recent reviews. The relatively small sizes of GMCs currently limit
detailed observations of such clouds to nearby galaxies, with recent
surveys done in the MW (\cite{solo87,heye01}), M33 (\cite{roso04}) and
the LMC (\cite{fuku01}).

\cite{blit06b} present detailed observations of GMCs in five local
galaxies. GMCs are excellent tracers of spiral structure to the extent
that they are often used to {\em define} the location of arms, in a
similar way as, for example, H{\sc II} regions or massive stars
are. There is a good correlation between the locations of GMCs and
filaments of H{\sc I}, although H{\sc I} is often observed without
accompanying molecular clouds, suggesting that the clouds form from
H{\sc I}.

Observations in more distant galaxies are currently limited to
measuring the mean surface density of molecular gas with more detailed
observations awaiting new instruments such as
ALMA\footnote{http://www.alma.nrao.edu/}. According to \cite{youn91},
the fraction of gas in the molecular phase depends on Hubble type,
with early-type spirals tending to be predominantly molecular and
late-types atomic. Optically barred spirals show a clear enhancement
of CO emission along the bar. This suggests that the large-scale
structure of the galaxy affects the formation of clouds and hence,
presumably, also the star formation rate. A physical implementation of
star formation should therefore aim to track the formation and
evolution of molecular clouds. But how do GMCs form?

\subsection{The formation of molecular clouds}
Although the early models by \cite{fiel69} assumed that the different
phases of the ISM were in pressure equilibrium, modern observations
paint a picture of a much more complex and dynamic situation in which
the ISM is shaped by turbulence, possibly powered by SNe and the
large-scale dynamics of the galactic disk itself, see
e.g. \cite{burk06} for a recent review. A galaxy-wide effect seems to
be required to explain the observed Hubble-type dependence of cloud
properties.

Yet how GMCs form in this complex environment is not well understood
(\cite{blit04}). Some authors have suggested that GMCs form by the
coagulation of preexisting molecular clouds (\cite{kwan83,oort54}) while
others have argued that GMCs form primarily from atomic gas through
instability or large-scale shocks (\cite{blit80a}). A viable mechanism
by which this could occur is the formation of convergent flows induced
by a passing spiral arm (\cite{ball99}). Of course, both modes of
formation may occur: in high density regions, where the vast majority
of hydrogen is molecular it seems likely that molecular clouds form
from the coagulation of smaller clouds, whereas in the outskirts of
galaxies where the gas is predominantly atomic the compression of gas
in spiral density shock waves provides a more plausible formation
mechanism.

Observationally, star formation in disks seems to occur only above a
given surface-density threshold (\cite{kenn89}), with star formation
dropping abruptly below the threshold even though the gaseous disk may
extend far beyond it. \cite{scha04} describes a model in which this
threshold arises naturally due to the thermal instability when gas
cools from $10^4$K to the cold phase ($\sim 100$K), rendering the
disk gravitationally unstable on a range of scales. This suggests that
a combination of thermal instability and large-scale dynamics may be
responsible for determining when and where GMCs form.

\cite{elme00} discusses observational evidence that GMCs are in fact
short lived entities that form, make stars and disperse again on their
dynamical time scale. This short time-scale alleviates the need for an
internal energy source to sustain the observed internal supersonic
turbulence, something that had puzzled astronomers for a long
time. \cite{prin01} discuss this assumption in more detail and
suggest that GMCs form from agglomeration of the dense phase of the
ISM, already in molecular form, when compressed in a spiral
shock. They envisage the pre-existing molecular gas to be in dense \lq
wisps\rq, in the inter arm regions, formed from atomic gas by shocks,
as simulated by \cite{koya00}.

Recent numerical simulations support the view that when clumpy gas is
overrun by a density wave it produces structures that resemble
GMCs. \cite{wada02} present high-resolution two-dimensional
simulations of the evolution of perturbations in a cooling,
self-gravitating disk in differential rotation. They show that the
disk develops stationary turbulence, even without any stellar
feedback. \cite{bonn06} and \cite{dobb06} performed three dimensional
simulations of the passage of clumpy cold gas through a spiral
shock. Their simulations produce dense clouds, with large internal
velocity dispersion, reminiscent of the \lq supersonic turbulence\rq\,
seen in GMCs. They note that the velocity dispersion is generated on
all scales simultaneously, in contrast to what is usually meant by
turbulence where energy cascades from large to small
scales. \cite{macl04} review the current state of the art in
simulations and models of GMC formation, including references to more
recent work. In this picture of clouds, GMCs are temporary structures
formed and dissolving in converging flows. They do not require an internal
source of energy, are not in virial or pressure equilibrium and need
not even be gravitationally bound. They point out that the relative
contribution of galactic rotation and stellar sources to driving the
observed turbulence is not clear.

\subsection{Star formation in molecular clouds}
In the older picture of cloud formation, GMCs were long-lived,
gravitationally bound, virialised objects. The presence of supersonic
turbulence ensures that clouds do not immediately collapse to form
stars, as this would predict a star formation rate for the MW which is
far higher than observed. Locally unstable clumps collapse to form
proto-stars, which built-up their mass to produce the initial mass
function (IMF) through competitive accretion (e.g. \cite{bonn97}).

However, simulations show that the energy contained in supersonic
motions is quickly dissipated even in the presence of magnetic fields
(see references in \cite{macl04}). To sustain the turbulence therefore
requires an energy source, for example from star formation, yet some
clouds have turbulence but no current star formation.

The modern picture is one in which clouds are short-lived structures
and the turbulence results from the same process that formed the cloud
in the first place. Observationally, GMCs turn a small fraction
$\epsilon_\star \approx 0.1$ of their mass into stars before they disperse
again. This low star formation efficiency of clouds may be due to the
fact that they are short lived. The short life times of (star forming)
clouds also follows from the small age spread in star clusters (see
e.g. \cite{gome92}), and indicates that star formation in a given
cloud only lasts for a few million years. The short life-times of GMCs
then also suggests that competitive accretion (e.g. \cite{bonn97}) is
less important in shaping the IMF (\cite{pado02}).

Turbulence generates a range of substructures inside a GMC, and
\cite{pado02} suggest that such \lq turbulent fragmentation\rq\ builds
a mass spectrum of proto-cores, some of which will collapse under
their own gravity to form stars. The resulting IMF is a power-law due
to the self-similar nature of the turbulence. Only cores dense enough
so that self gravity can overcome their magnetic and thermal energy
can collapse. This consideration flattens that IMF at low masses, and
prevents very low-mass cores from forming stars. They also argue that
the maximum mass is a fraction of the overall cloud mass.

A young stellar population does, of course, dump a lot of energy into
its surroundings through stellar winds, ionisation, and SN
explosions. Even if these do not drive the observed turbulence, they
may contribute to the destruction of the cloud, and prevent further
star formation. Most simulations use such feedback from star formation
to regulate the star formation rate.

\subsection{Summary}
The current theory of star formation in disk galaxies suggests that
supersonic turbulence, generated by a combination of galactic rotation
and SNe, regulates the formation of proto-stellar cores inside more
massive molecular clouds. Some fraction of these clouds can form
stars, before the cloud itself is destroyed, by a combination of
stellar feedback and the turbulence that built the cloud in the first
place. The cooling of the cores to temperatures $\le 100$K is
dominated by grains and CII fine-structures lines and is opposed by
photo-electric heating from small grains and polycyclic aromatic
hydrocarbons (PAHs), see \cite{wolf95}.

A numerical implementation of these processes requires high resolution
to model the interstellar turbulence and follow the contraction of
cores of masses a few $M_\odot$ at densities above 1 particle per
cm$^{-3}$.  Such challenging simulations may be feasible in the near
future for high-redshift galaxies but are not possible yet for $z=0$
galaxies. Below we describe a model of cloud formation that tries to
incorporate some of these processes with some simple rules.

We would like to apply the same rules to cosmological simulations of
galaxy formation, which clearly requires a leap of
faith. High-redshift galaxies may not have a well-defined disk, and
hence the properties of the supersonic turbulence and the GMCs may
well be very different. In the high-redshift simulations of
\cite{abel00}, the first molecular cloud is close to hydrostatic
equilibrium, with pressure support slowly leaking away as it cools
through molecular hydrogen line emission. This quasi-static evolution,
reminiscent of a cooling flow, is very different from the dynamic
turbulent fragmentation envisaged in the MW, with corresponding large
differences in the predicted IMF. Furthermore, if the properties of
the GMCs were similar, the behaviour of the cores may still be very
different, with the reduced grain and metal cooling at higher $z$,
making for a different IMF.

Despite these difficulties we must start somewhere.  The assumption
that the physics of the ISM (and therefore the stellar IMF) is similar
at redshift zero and in the high redshift universe is common in the
simulation community.  The understanding gained through these
necessarily simplified simulations will allow us, over time, to
investigate the physics relevant to galaxy formation at higher
redshift.
 
\section{Details of the Model}
\label{sec:model}

As demonstrated in section \ref{sec:intro}, in typical simulations of
galaxy formation we can resolve the Jeans length of the ambient gas
phase and so treat its hydrodynamic properties consistently.  However,
we cannot yet resolve the properties of the cold molecular phase of
the ISM.  We therefore follow the evolution of the ambient gas phase
using a hydrodynamic simulation code, whereas we treat the cold phase
using a statistical model that encapsulates the physics relevant to
the formation and evolution of molecular clouds. In this section we
introduce the properties of the sticky particle model and describe the
physics we have implemented.

Following \cite{efst00} we consider the ISM to consist of warm and
hot ambient materials, and cold molecular clouds.  We additionally
treat the properties of SN remnants.  Throughout this paper the
properties of the ambient medium will be represented with the
subscript $h$, the properties of the molecular clouds with the
subscript $c$, and the properties of the gas internal to SN remnants,
or hot bubbles, with the subscript $b$.

The ambient gas phase is represented using the entropy conserving,
parallel Tree-SPH code GADGET2 (\cite{spri05,spri01}), which is a
Lagrangian code used to calculate gravitational and hydrodynamic
forces on a particle by particle basis.  Smoothed-Particle
Hydrodynamics (SPH) was originally introduced by \cite{lucy77} and
\cite{ging77}, see e.g. \cite{mona92} for a review.  We will refer to
the gas component treated using SPH as ambient gas to distinguish it
from the cold molecular gas.  Ambient gas at temperatures around
$10^4K$ will be referred to as \lq warm\rq\,, and gas at temperatures
of $10^6K$ and higher will be called \lq hot\rq\,. We will see that in
galaxy formation simulations, this ambient (i.e. non-molecular) medium
naturally develops three relatively well-defined phases: a warm
($T\sim 10^4$K) component in a galactic disk, a hot ($T\sim 10^6$K))
tenuous component of shock-heated gas in the halo, and a similarly hot
component resulting from gas heated by SN. The fourth, cold ($T\sim
100$K) and molecular cloud phase is represented with sticky particles,
which interact gravitationally with all other material in the
simulation and are allowed to stick together forming more massive
sticky particles.  Stars and dark matter are both treated as
collisionless particles by GADGET2.

The different phases of the ISM may interact with each other as
follows: thermally unstable ambient gas may collapse into molecular
clouds via thermal instability (section \ref{sec:cloud-form}).
Molecular clouds can interact with each other to form GMCs (section
\ref{sec:coagulation}).  GMCs then collapse into stars (section
\ref{sec:cloud-collapse}). Stars disrupt the cloud they formed from
and may, via SN feedback, return energy (section
\ref{sec:energy-feedback}) to the ambient phase.  Hot bubbles blown by SNe
can evaporate cold clouds (\ref{sec:conduction}) and heat the ambient
medium.

\begin{figure}
\begin{center}
\includegraphics[width=8.3cm,clip]{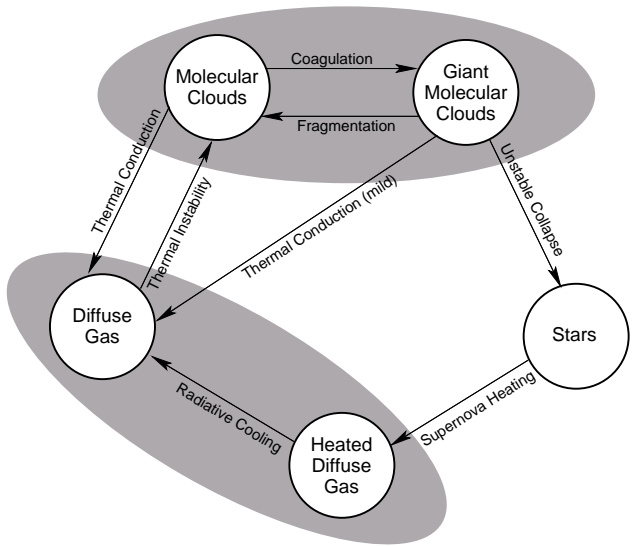}
\end{center}
\caption{Summary of the physical processes that operate in our model
of a two phase interstellar medium.  The boundaries between molecular
and giant molecular clouds and between heated and non-heated diffuse
gas are somewhat arbitrary and they are separate out in this figure
only to highlight the different physical mechanisms that are operating
at any given time.}
\label{fig:modsumm}
\end{figure}

Fig \ref{fig:modsumm} contains a summary of all of the physics
implemented in our model. Arrows represent a transfer of mass and/or
energy from one phase to another.  The distinction between clouds and
GMCs is somewhat arbitrary; they are separated in the figure to allow
an easy pictorial representation of mass transfer within a single
phase. Appendix \ref{app:list-of-symbols} contains a list of
frequently used symbols and their meaning.

Each physical process will be treated in turn in the remainder of this
section. We first introduce the physics relevant to each physical
process before discussing the numerical implementation. We will also
give our preferred physical values for the various parameters that
occur. How we choose these is discussed in section ~\ref{sec:parameter-est}.

\subsection{Radiative Cooling And The Formation of Molecular Clouds}
\label{sec:cloud-form}

\cite{bege90} show that under appropriate physical conditions, a
thermal instability may operate in the ambient gas, which causes a
fraction of the gas to condense into much colder molecular clouds. The
sticky particle star formation prescription contains a basic
representation of this process, based on a detailed treatment of
baryonic radiative cooling.

\subsubsection{Relevant Physics}
The radiative processes that we take into account are Compton cooling
off the microwave background, thermal Bremsstrahlung cooling, line
cooling and photo-ionization heating from Hydrogen, Helium and metal
species in the presence of an imposed ionising background.  These
routines were developed for a different project and will be described
elsewhere\footnote{We would like to thank our colleagues J Schaye, C
Dalla Vecchia and R Wiersma for allowing us to use these rates.}.
Briefly, they use tabulated rates for radiative cooling and
photo-ionization heating for many species and ionization states
computed assuming ionization equilibrium using {\sc cloudy} (version
05.07 of the code last described by \cite{ferl98}) with a UV
background given by \cite{haar01}. The rates are tabulated element by
element and we will assume solar abundance ratios and specify a fixed
metallicity of the gas in solar units.  We do, however, note that the
behaviour of the system may depend upon precisely which value of the
metallicity we choose, and investigate this in section \ref{sec:away}

Other processes such as cosmic ray heating, and cooling by dust and
atomic lines that affect the molecular gas in clouds are not treated
explicitly since we do not model the internal properties of the clouds
themselves.

We include a simple model to determine the rate at which the ambient
gas forms molecular clouds. When we identify ambient gas that is
thermally unstable (\cite{bege90}) we allow it to collapse into
molecular clouds.  The rate at which this process occurs is goverened
by the rate at which the gas is losing thermal energy by radiative
cooling.

\subsubsection{Numerical Implementation}

Following \cite{yepe97} we define a density threshold, $\rho_{\rm
th}$, to determine when gas becomes thermally unstable.  Gas with
$\rho< \rho_{{\rm th}}$ undergoes ordinary radiative cooling. Gas with
density above the threshold becomes thermally unstable and begins to
be converted to molecular clouds.  In addition to the density
criterion we add a maximum temperature ($T_{{\rm th}}$) for gas to be
called thermally unstable which has the effect of preventing SN heated
gas in dense regions from collapsing straight to the cold phase.

When gas has been identified as thermally unstable it begins to form
molecular clouds at a rate controlled by the rate at which the gas can
lose thermal energy by radiative cooling
\begin{equation}
\frac{d\rho_{c}}{dt} = -\frac{d\rho_{h}}{dt} = \frac{1}{u_{h}-u_{c}}\Lambda_{{\rm net}}(\rho_{h},u_{h}),
\label{eq:coldform}
\end{equation}
where $u$ represent an internal energies per unit mass.  The
subscripts $h$ and $c$ refer to the ambient phase (either warm or hot)
and cold phase respectively. $\Lambda_{{\rm net}}$ is the cooling rate
of the ambient gas (ergs cm$^{-3}$ s$^{-1}$).  We assume that the cold
clouds remain at a fixed temperature of $T_{\rm c}=100$K hence their
thermal energy $u_{\rm c}$ is a constant as well.

In practice, each ambient gas particle is identified as either
thermally unstable, or non-thermally unstable.  non-thermally unstable
gas undergoes radiative cooling; thermally unstable ambient gas forms
molecular clouds at a rate controlled by the radiative cooling rate,
as described by Eq.~\ref{eq:coldform}.

In this way each ambient gas particle can keep track of what fraction
of its mass is in the form of molecular clouds.  Gas in the molecular
phase is ignored for the purposes of the SPH calculation.  When the
amount of mass in the molecular phase in a particle reaches the
resolution limit of the simulation a separate \lq sticky particle\rq,
representing many sub-resolution molecular clouds is created.  This
process decouples the molecular clouds from the associated ambient
phase.  Since we cannot resolve the individual molecular clouds in
each sticky particle we work with the mass function of clouds.
Initially we assume that the molecular clouds formed through
instability are all in the smallest mass bin, that is that the clouds
formed by thermal instability are very small, and will interact to
form more massive clouds.  In the following section we describe the
behaviour and evolution of the sticky particles in the simulation.

\subsection{Cloud Coagulation and GMC Formation}
\label{sec:coagulation}

Molecular clouds are typically many orders of magnitude more dense
than the medium they form in (MO77), and their behaviour is governed
by a different set of rules than the ambient medium.  This section
describes the physics of the simplified molecular clouds in the sticky
particle model and how it is implemented.

\subsubsection{Relevant Physics}
We assume that clouds may be treated as approximately spherical
objects that obey a power law relation between mass ($M_c$) and
radius ($r_c$) 
\begin{eqnarray}
\label{eq:r-m}
r_c &=& \Big( \frac{M_c}{M_{\rm ref}}
\Big)^{\alpha_{c}}r_{\rm ref}\nonumber\\
                       &=& 36\,\Big(\frac{M_c}{10^{5}
  M_{\odot}}\Big)^{0.3}\,{\rm pc}\,.
\label{eq:cloud_mass_radius}
\end{eqnarray}
Here, $\alpha_{c}$ describes how clouds grow as mass is added to them
(if they remain at constant density then $\alpha_{c}=1/3$), and
$M_{\rm ref}$ and $r_{\rm ref}$ are a reference mass and radius used
to fix the normalisation of this relation. The lower bound on
molecular cloud masses is typically calculated to be 100$M_{\odot}$
(\cite{mona04}) due to the efficient destruction of smaller molecular
clouds by photoionization. We introduce an upper limit by converting
molecular clouds with large masses into stars (see section
\ref{sec:cloud-collapse} for discussion).  In order to facilitate easy
estimates of the relative importance of various effects we have
substituted typical numbers and units into most of the equations in
this section.

\subsubsection{Numerical Implementation}

Each sticky particle represents numerous cold clouds.  Sticky
particles are hydrodynamically decoupled from the ambient SPH phase of
the gas and interact only gravitationally with the other phases in the
simulation.  However, when two sticky particles collide they may
coagulate to form a more massive sticky particle.  The mass of the
smallest molecular clouds is typically orders of magnitude below the
mass resolution in a cosmological simulation.  We represent an entire
mass spectrum of clouds statistically inside of each sticky particle.
Our formalism to treat the evolution of the mass function of clouds
internal to each of the 'multiple cloud' particles will start from the
Smoluchowski equation of kinetic aggregation (\cite{smol16}), which
describes the behaviour of a system consisting of ballistic particles
that can interact via mergers.  The coagulation behaviour of this
system is driven by a coagulation kernel, $K(m_1,m_2)$, that
represents the formation rate of clouds of masses $m=m_1+m_2$,

\begin{equation}
K=\langle \Sigma v_{{\rm app}}\rangle_{v},
\label{eq:k}
\end{equation}
where $v_{{\rm app}}$ is the relative velocity of the clouds and
$\Sigma$ is the collision cross section. For a Maxwellian distribution
of velocities with three-dimensional dispersion $\sigma$ we obtain
$\langle v_{{\rm app}}\rangle =1.3\sigma$ (\cite{lee88}). The product
of the approach velocity and the collision cross section is averaged
over the distribution of relative velocities.  The cross section is
\begin{equation}
\Sigma \approx \pi(r_c+r_c')^{2}\Big(1+2G\frac{M_c+M_c'}{r_c+r_c'}\frac{1}{v_{{\rm app}}^{2}}\Big)\,,
\label{eq:sasl}
\end{equation}
where the first term represents the collision geometric cross section
and the second term represents the effect of gravitational focusing
(\cite{sasl85}).  The focused term becomes significant when the
approach velocity is not much larger than the internal velocity
dispersion of the system.  In most cases of interest the geometric
term will dominate so the focused term is neglected.  In these
calculations we need to assume that molecular clouds, although
transient and turbulent, are stable for long enough for coagulation to
take place.  This is reasonable because the cloud velocity dispersion
is typically larger than the sound speed of the cold cloud gas
(\cite{mona04}).

To model the cooling of sub-resolution molecular clouds via
gravitational interaction it has been assumed that when molecular
clouds with relative velocities, $v_{{\rm app}}$ greater than $v_{{\rm
stick}}$ (a parameter in our simulations) collide they do not merge,
but rather bounce back with relative velocity a fraction, $\eta$, of
the initial approach velocity. Clouds with relative velocities less
than $v_{{\rm stick}}$ merge.  For simplicity it has been assumed that
the velocity distribution of clouds is Maxwellian with a velocity
dispersion that is a function of cloud mass, $\sigma = \sigma(m)$.

The upper and lower bounds on the molecular cloud mass function are
set such that the smallest mass bin is comparable with the smallest
clouds was can observe, and the largest molecular clouds are
approximately the same mass as the largest clouds in the MW.  The mass
function is discrete.  All clouds are assumed to form at the lowest
mass, $M_{min}$, and then the mass of each bin is a multiple of this
value.  This discrete mass function is neccessary when working with
the Smoluchowski equation.

In order for us to be able to hold a mass function with a large number
of bins internal to every single sticky particle without the
requirement to store one number for each mass bin we parameterize the
mass function as a third order polynomial, and store only the four
coefficients between timesteps.

As these sub-resolution clouds interact and merge, the one-dimensional
velocity dispersion $\sigma(m)$ changes, which affects the rate of
evolution of the cloud mass function, $n(m)$. Let $E_{\rm m}=3/2
m\sigma^2(m)$ denote the random kinetic energy of clouds with mass
$m$. The one-dimensional velocity dispersion is related to the three
dimensional velocity dispersion by
$\sigma_{1D}=\sigma_{3D}/\sqrt{3}$. $E_{{\rm m}}$ and may change due
to three distinct processes:
\begin{itemize}
\item Clouds with masses $m'$ and $m-m'$ merge to form extra
  clouds of mass $m$, increasing $E_{{\rm m}}$ at a rate $\dot{
  E}_{{\rm gain}}$
\item Clouds with masses $m$ may merge with clouds of any other mass
  decreasing the number of clouds of mass $m$.  This process decreases
  $E_{{\rm m}}$ at a rate $\dot{E}_{{\rm loss}}$
\item Clouds with mass $m$ may interact collisionally with clouds of
  any other mass and so lose kinetic energy.  This process decreases
  $E_{{\rm m}}$ at a rate $\dot{E}_{{\rm cool}}$
\end{itemize}

The net change in kinetic energy for particles of mass $m$ during some
timestep $\Delta t$ is given by
\begin{equation}
\label{eqn-ke}
\Delta E_{m} = \frac{dE_{m}}{dt}\Delta t = 
\Big[\dot{E}_{{\rm gain}}
- \dot{E}_{{\rm loss}} - \dot{E}_{{\rm cool}}\Big]\Delta t\,.
\end{equation}
And for this change in kinetic energy, the corresponding change in velocity dispersion is given by
\begin{equation}
\dot{\sigma}=\frac{2\dot{E}-\dot{M}\sigma^2}{2M\sigma}\,.
\label{eqn-sigma}
\end{equation}
 Details of the equations used to model these processes are
given in Appendix \ref{app:co-derive}, and the method by which they
are solved numerically in Appendix \ref{app:co-solve}.

The same processes (cooling and merging) are followed explicitly for
the individual particles in our simulations, which can interact in the
same two ways as the unresolved sub resolution clouds by merging to
form more massive sticky particles, or cooling to decrease their
relative velocities.  Following the same rules should allow us to
remove much of the resolution dependence of the star formation
prescription.  As the mass resolution of a simulation is degraded,
more massive clouds will be treated with the sub-grid physics; our
implementation should ensure that the large scale results are
approximately the same.  This is demonstrated in section
\ref{sec:calib}.

\subsection{Cloud Collapse and Star Formation}
\label{sec:cloud-collapse}

The vast majority of stars form in Giant Molecular Clouds. This
process is described in the sticky particle model by allowing the most
massive clouds in the galaxy to collapse into stars.

\subsubsection{Relevant Physics}

We follow the process of star formation in our simulations by waiting
for star forming clouds to be created by the coagulation process
described in section \ref{sec:coagulation}.  We define star forming
clouds to be clouds of a mass similar to the most massive clouds
observed in the MW ($\sim 10^6 M_{\odot}$).  When one of these star
forming clouds is created it is assumed to collapse on a short
timescale and approximately $\epsilon_\star\sim 10$\% of its mass is
converted into stars, whilst the remainder is disrupted by stellar
feedback processes including stellar winds, SN feedback and
photoionization.  This process reflects that although stars may form
in less massive molecular clouds, it is not until the relatively rare,
massive O and B stars are created that the cloud is destroyed
(\cite{elme83}).

We assume that each cloud collapse forms a single stellar population
with an IMF of the standard \cite{salp55} form
\begin{equation}
\label{eq:imf}
N(M)\,dM \propto M^{-(1+x)}dM\,,
\end{equation}
where $x$ is the slope of the IMF and takes the usual value of 1.35.
The masses of stars are assumed to lie between well defined minimum
and maximum values, $M_{{\rm \star,min}}$ and $M_{{\rm \star,max}}$.

\subsubsection{Numerical Implementation}
The treatment of star formation adopted in most simulations is to
identify gas that is likely to be star-forming and impose a
star formation rate given by the Schmidt law,
\begin{equation}
\dot \rho_{{\star}}=C\rho_{{\rm gas}}^{N_{{\rm SF}}}\,.
\label{eq:schmidtlaw}
\end{equation}
Here, $\dot\rho_{\star}$ and $\rho_{\rm gas}$ denote the rate of star
formation per unit volume and the gas density respectively.  This
power law relation between star formation rate (SFR) and gas density
was found to hold over many orders of magnitude by \cite{kenn98}, who
constrained the exponent to be $N_{\rm SF}=1.4\pm0.2$.

We take a different approach: unstable molecular clouds are identified
in the simulations as any cloud with a mass greater than $M_{{\rm
sf}}$.  We identify the formation of these massive clouds by using the
cloud mass function, as stored internally to every single sticky
particle.  These unstable clouds are assumed to collapse on a very
short timescale, forming stars.

As soon as a cloud of mass $M_{{\rm sf}}$ forms, it is assumed to be
disrupted by OB stars on a timescale of $\sim $10Myr (\cite{matz02}),
the rest of the massive cloud is broken down into smaller clouds and
the coagulation process begins all over again as described in section
\ref{sec:coagulation}.  This process is modelled by taking the
fraction of the cloud's mass that does not turn into stars,
$1-\epsilon_*$, and assuming that the net effect of the stellar
feedback processes is to fragment the GMC into the smallest clouds
represented in the sticky particle internal mass function.  This has
the net effect of steepening the cloud mass function.

Each star particle formation event represents the formation of a
single stellar population of stars that are all assumed to have the
same age, and to be drawn from the Salpeter IMF.  Each stellar
particle is therefore formed with a mass approximately equal to
$\epsilon_*$ times the mass of a starforming cloud.  If this particle
mass is not allowed by the mass resolution of a given simulation then
we either store up unresolved stars internal to a sticky particle (if
the star mass is too small to be allowed), or split it into multiple,
equal mass particles (if the star mass is too large to be allowed).

\subsection{Supernova Feedback}
Our simulations include only energy feedback from type II SN.  These
events return energy from the stars to the ambient phase.  We note
that it is not currently computationally feasible to resolve the
properties of SN remnants so we treat them with a simple, analytic
prescription.  The mechanism by which SN feedback is implemented in
our model is discussed here.

\subsubsection{Relevant Physics}
\label{sec:energy-feedback}
Each star of mass greater than $8M_{\odot}$ releases $10^{51}E_{51}$
ergs in thermal energy when it undergoes an SN event.  The lifetime,
$t$, of a star of mass $M$ (where $M>6.6M_{\odot}$) is given by
(\cite{pado93})
\begin{equation}
\label{eq:stellarlifetime}
\frac{t}{{\rm Gyr}}=1.2\,\Big(\frac{M}{M_\odot}\Big)^{-1.85}+0.003\,.
\end{equation}

Each SN explosion can be approximated as the injection of energy at a
single point in space. If we assume that the ambient density on scales
of interest is approximately homogeneous, with density $\rho_h$, then
each SN explosion can be modelled as a Sedov blast wave
(\cite{sedo59}). According to this solution, if at time $t=0$ we
release an amount of energy $E_b$, then after time $t$ the resulting
blast wave will have reached a radius $r_b$ given by
\begin{eqnarray}
r_{b}&=&\Big(\frac{E_b}{\rho_{h}}\Big)^{1/5}t^{2/5}\nonumber\\
           &=& 292\, \Big(\frac{E_b/10^{51}{\rm ergs}}{\rho_{h}/0.1\,{\rm cm}^{-3}}\Big)^{1/5}(t/10{\rm
  Myr})^{2/5}\,{\rm pc}\,.
\label{eq:sedov}
\end{eqnarray}

These hot SN bubbles have two main effects. Firstly, as they expand and
decelerate the SN heated gas will get mixed in with the surrounding
ambient medium; the net result of this process is the heating of the
ambient medium.  Secondly, as discussed in section \ref{sec:conduction},
any cold clouds caught inside an SN bubble will undergo
evaporation.

There are two main assumptions that must hold for the Sedov solution
to be valid, the pressure of the ambient medium, and the cooling rate
inside the bubble, must both be negligible. Often at least one these
assumptions is invalid. If the ambient medium has a low density and is
very hot, for example due to a previous set of explosions, then its
pressure is no longer negligible and the Sedov solution breaks down.
If the ambient medium is dense then radiative cooling becomes an
important process.  In the remainder of this section we describe
various modifications to the standard Sedov solutions, which allow us
to model SN remnants in a wider variety of conditions.

\begin{figure}
\begin{center}
\includegraphics[width=8.3cm,clip]{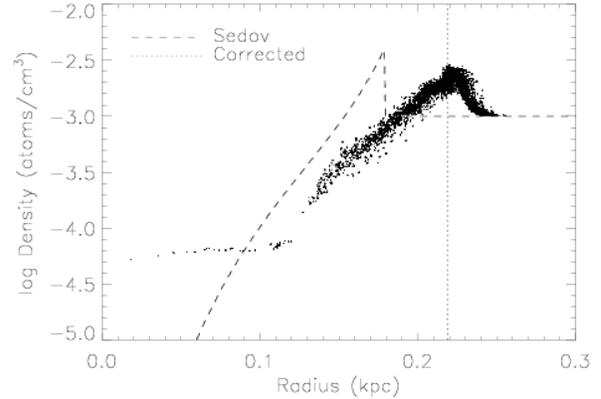}
\end{center}
\caption{Comparison between SPH simulation of a Sedov blast, the Sedov
solution and the hot medium correction of Tang \& Wang (2005).  The
points represent individual SPH particles, the dashed line is the
Sedov solution and the dotted line is the blast wave radius as
calculated with the hot medium correction.  The initial condition had
a density of 0.001 atoms per $cm^3$ and a temperature of $10^6$K.
$10^{51}$ergs were injected to the central 32 particles at $t_{0}$.
This plot was made after the blast wave had evolved for 0.3 Myr.}
\label{fig:sedov}
\end{figure}

In the case of a hot, tenuous medium the radius of each blast wave is
increased (\cite{tang05}). These authors derive a fitting formula for
the velocity of a SN blast in a hot medium, which is accurate
to within 3\%
\begin{eqnarray}
\label{eq:tang}
r_{b}(t) &=&\int^t_0 c_{h}\Big(\frac{t_c}{t'}+1\Big)^{3/5}dt'\,,\\
	 &=&156 \int^{t/{\rm Myr}}_0 \Big(\frac{t_c}{t'}+1\Big)^{3/5}dt'\; {\rm pc},
\end{eqnarray}
where $c_{{\rm h}}$ is the sound speed of the ambient medium. We
assumed a temperature of $T_h=10^6$K, mean molecular weight of
$\mu=0.58$, blast wave energy of $1\times 10^{51}$ ergs and an ambient
density of $0.1$ atoms per cm$^3$ in order to illustrate the order of
magnitude of $r_b$.  $t_{c}$ is a characteristic time,
\begin{eqnarray}
\label{eq-tc}
t_{c} &=&\Bigg[\Big(\frac{2}{5}\xi\Big)^5\frac{E_b}{\rho_{h}c_{
		  h}^5}\Bigg]^{1/3}\,\nonumber\\
      &=&0.012\,\Bigg[(\xi/1.14)^5\,{E_b/10^{51}{\rm erg}\over
		  (\rho_h/0.1\,{\rm cm}^{-3})(T_h/10^6{\rm
		  K})^{5/2}}\Bigg]^{1/3}\,{\rm Myr}\nonumber\,.\\
\end{eqnarray}
where $\xi$ equals 1.14 for a gas with adiabatic index $\gamma=5/3$.
This solution matches the standard Sedov evolution, $r_b\propto
t^{2/5}$, closely until $t\sim t_c$, after which the shell's velocity
becomes constant, $r_b\propto t$. This modification allows us to take
into account that a the majority ($\sim$90\%), of SNII happen in preheated
SN bubbles (\cite{ higd98}) and, therefore, the approximation that the
pressure of the ambient medium is negligible is often incorrect.  Fig
~(\ref{fig:sedov}) shows the difference between an adiabatic gas SPH
simulation of a SN induced shock-wave, the pure Sedov solution and
the blast wave radius as predicted by the hot medium-modified Sedov
solution from Tang \& Wang (2005).

Situations where radiative cooling are important may be taken into
account using the prescription of \cite{thor98}, whose high resolution
simulations of SN explosions expanding in an ambient medium with
temperature $T_h=10^3$K, provide the total thermal energy in SN bubbles
as a function of time, ambient density and metallicity. We perform
bilinear interpolation on the results in tables 2 and 4 of
\cite{thor98} to obtain the SN bubble radius and thermal energy
at any given time.

Neither of these solutions treats the more general case of SN remnant
expansion in a porous ISM, which may have regions of both high and low
ambient density, and so we are not able to include the effects of SNe
fully self-consistent. In most of our simulations we use the simple
Sedov solution for the evolution of the SN blast waves, but note that
the details of our prescription are uncertain.  In section
\ref{sec:away} we investigate the effects of using different
implementations of the blast wave implementation to estimate how
important the details of the behaviour of SN remnants are to the
overall properties of the galaxy. Both the radiative cooling and blast
wave velocity physics are varied.

\subsubsection{Numerical Implementation}
\label{sec:snnumerics}
By assuming that each stellar particle in the simulation represents an
entire population with the same age we can calculate the minimum and
maximum masses of stars that undergo supernova events over any given
time period using Eq ~(\ref{eq:stellarlifetime}).  Each of these
supernovae is assumed to go off in a neighbouring gas particle ({\em
i.e.} one for which the distance, $r$, to the star is smaller than its
smoothing length, $h$, in the SPH formalism). We chose this particle
randomly from the neighbours, with a weight computed from the solid
angle, $\Omega$, it subtends on the sky as seen from the position of
the star particle,
\begin{equation}
\Omega=4\pi\,\Big(1-{r\over\sqrt(r^2+h^2)}\Big)\,.
\end{equation}
This weighting forces that nearby hot, diffuse gas (which tends to have
larger $h$, hence larger weight) is heated more frequently than cooler, denser parts of the ambient medium (which are dense, hence have smaller $h$).

We do not transfer all SN energy to gas particles each timestep.
Assuming that SN explosions are distributed evenly in time and space
we can calculate for every ambient gas particle a \lq porosity\rq\, of
SN bubbles, $Q=V_{{\rm B}}/V_{{\rm A}}$. For the volume associated with
a gas particle we use $V_{{\rm A}}=(4\pi/3)\,h^3$, and $V_{{\rm
B}}=(4\pi/3)\sum r_b^3$ is the total volume of all the SN bubbles in
this particle. When $Q$ is greater than a critical value, $Q_{\rm
crit}\approx 1$, the ambient phase is heated, else the available SN
energy is carried over to the next time step. This ensures that the
ambient phase is only heated when hot supernova bubbles make up a
significant fraction of the volume. There are two motivations for this,
firstly a given SPH particle cannot represent more than one phase at a
given time. Secondly simulations usually do not limit the timestep to
be a fraction of the cooling time. Consider a warm, $T\sim 10^4$K, SPH
particle in the disk. If a small amount of SN energy is injected into
this dense particle, it will cool very efficiently since the cooling
rate is very high. It is only when the particle is heated to $T\gg
10^6$K that the reduced cooling may affect the particle dynamically, so
that it will move into lower density gas, further decreasing its
cooling rate, and becoming part of the hot, tenuous gas. Storing the
available heating until the SN bubbles fill a significant fraction of the
particle is a way of easing the transition from warm to hot and makes
the outcome less dependent on the timestep.

To determine the porosity $Q$, we need to know the current radii,
$r_b$, of SN bubbles. The radius $r_b$ depends on the ambient gas
properties and also on the available energy, $E_b$, as discussed in
section \ref{sec:energy-feedback}. Typically a single stellar particle
will undergo multiple SN events over a single timestep.  Using
Eqs.~(\ref{eq:stellarlifetime}) and (\ref{eq:sedov}) and obtaining the
SPH estimate of the ambient gas density at the position of the star
particle we can estimate the average radius of all supernova bubbles
blown by a given star particle at any time.  Working under the
assumption that the porosity of the ISM is low we calculate the
radiative loss from each bubble separately. When the porosity of the
ISM becomes $Q>Q_{\rm crit}\sim 1$, the SN bubbles are overlapping
significantly and all coherent structure is assumed to be wiped out.
The ambient gas particles are heated by the remaining thermal energy in
the supernova bubbles and they are considered to disperse. The porosity
is set back to zero. Note that using the Sedov solution implies we
neglect radiative cooling in the remnants to determine the porosity,
$Q$.  However to determine how much energy is in the bubbles once we
decide to heat the particle, we do use the tables of \cite{thor98} to
account for radiative cooling in the SN shells. We believe that even
though this treatment is not fully consistent, it does capture the main
physics.

\subsection{Thermal Conduction}
\label{sec:conduction}

Thermal conduction between the ambient and cold gas in the simulation
is an important ingredient in the self-regulation of the star
formation rate in our model of the ISM.

\subsubsection{Relevant Physics}
\label{sec:thermphys}
Thermal conduction has two primary effects. The first is to smooth out
the temperature and density profiles inside SN remnants. In the strong
explosion solution of Sedov, where thermal conduction is neglected, the
temperature of the blast wave increases sharply towards the centre of
the blast.  This is due to the fact that the gas near the origin was
heated by a stronger shock than that at the edges and thereafter
evolves adiabatically.  The effect of thermal conduction is to
efficiently transport heat from the centre of the blast to the
outer cool regions.  The temperature of the interior of the supernova
blast, $T_{b}$, is then approximately constant and equal to the mean
temperature of the blast(\cite{chev75}; MO77):
\begin{equation}
\label{eq:Tf}
\Big(\frac{T_{b}}{{\rm 10^8K}}\Big)=
1.2\,
\Big(\frac{r_{b}}{{\rm 10pc}}\Big)^{-3}\,
(\frac{n_b}{0.1\,{\rm cm}^{-3}})^{-1}
(\frac{E_b}{10^{51}{\rm erg}})\,,
\end{equation}
where $n_b$ and $T_b$ are the mean density and temperature inside the
bubble, respectively. We assume $r_b$ to be described by Sedov's
self-similar solution. The density $n_b$ is also approximately constant
and is given in terms of the ambient density, $n_h$, as
\begin{eqnarray}
\label{eq:nh}
\frac{n_{b}}{n_{h}}&=&1+x^{-5/3}\\
x&\equiv &0.65\Big(\frac{r_b}{{\rm 10pc}}\Big)\Sigma_{{\rm
	 con}}^{1/5}(\frac{n_{h}}{{\rm
	 cm}^{-3}})^{3/5}(\frac{E_b}{10^{51}{\rm erg}})^{-2/5}\,.
\label{eq:x}
\end{eqnarray}
The dimensionless number $\Sigma_{{\rm con}}$ represents the
	 effectiveness of evaporation,
\begin{equation}
\Sigma_{\rm con}=\frac{\alpha_{\rm con}}{3} \Big(\frac{r_c}{{\rm pc}}\Big)^2 f_{{\rm cl}}^{-1}\phi^{-1},
\label{eq:sigmacond}
\end{equation}
(\cite{mcke77b}), and depends on $\alpha_{\rm con}=\dot r_b/c_h$ (the
ratio of the velocity of the supernova blast wave to the sound speed
of the medium), the cloud's radius, $r_c$, the volume filling factor
of the cold clouds, $f_{\rm cl}$, and the efficiency of thermal
conduction, $\phi$ (see MO77 for details). For a pure Sedov blast wave
$\alpha_{{\rm con}}=1.68$. The presence of magnetic fields and
turbulence may decrease $\phi$ below its maximum value of $\phi=1$.
We compute $f_{cl}$ for each sticky particle from its current cloud
mass spectrum given the assumed cloud mass-radius relation,
Eq.~(\ref{eq:cloud_mass_radius}).

The second effect of thermal conduction is to evaporate cold
clouds. According to (\cite{mcke77b,cowi77}), the evaporation rate is well described by:

\begin{equation}
\label{eq:Mdot}
\Big(\frac{\dot{M}_c}{{\rm M_\odot Myr}^{-1}}\Big)=-0.44\times \Big(\frac{T}{10^6{\rm K}}\Big)^{5/2}\Big(\frac{r_c}{{\rm pc}}\Big)\,.
\end{equation}

\subsubsection{Numerical Implementation}

Since we store the mass function of molecular clouds internal to each
sticky particle explicitly (Sect.~\ref{sec:coagulation}), we can apply
Eq.~(\ref{eq:Mdot}) along with Eq.~(\ref{eq:r-m}) to each cloud mass
bin to calculate the total mass loss of a cloud over one timestep.
The evaporation rate of the cloud depends on the temperature of the
ambient gas, which is represented with SPH particles. However, as we
discussed above, some fraction $Q$ of the volume of each SPH particle
may be filled by hot SN bubbles, in which the evaporation rate of
clouds may be much higher. Since we have computed $Q$, we can take
this important effect into account.

Consider a single molecular cloud in thermal contact with an ambient
medium of (constant) temperature $T$. The mass of a cloud at the end
of a timestep ($M_f$) is related to its mass at the start of the
timestep ($M_i$) by:
\begin{equation}
\label{eq:tc}
M_f=\Big[ M_{i}^{1-\alpha_c} - (1-\alpha_c)\frac{0.44T^{5/2}r_{{\rm ref}}}{M_{{\rm ref}}^{\alpha_c}}\Delta t \Big]^{1/(1-\alpha_c)}\,,
\end{equation}
where $T$ is in units of $10^6$K, masses are in $M_{\odot}$, lengths are in pc and times are in Myr.

Eq ~(\ref{eq:tc}) represents the mass loss rate for a single cloud in
contact with a medium of temperature $T$.  More generally in a porous
medium a single cloud of mass $m$ has a mean mass loss rate described
by:
\begin{equation}
\dot{M}_{\rm cloud}=-Q\dot{M}_{\rm bubble} - (1-Q)\dot{M}_{\rm ambient},
\end{equation}
where $\dot{m}_{\rm bubble}$ and $\dot{m}_{\rm ambient}$ represent the
rate of mass loss for a cloud inside a supernova bubble and situated
in the ambient medium respectively.  

Eq.~(\ref{eq:tc}) can be applied directly to the evaporation of a
cloud in the local ambient medium ($\dot{m}_{\rm ambient}$). However
to apply the same formula to the evaporation of clouds inside of
supernova bubbles we need to account for the fact that although the
temperature inside the bubbles remains uniform, due to conduction, it
is not constant in time, but decreases as the bubble expands. We
therefore make the additional assumption that the mean temperature of
the supernova remnant is constant over a timestep (a good
approximation after a short ($\sim 20$yr) transient phase).  Under
this assumption Eq.~(\ref{eq:tc}) can be applied successfully to the
more general case of evaporation in a porous medium.  Eq
~(\ref{eq:Mdot}) and Eq ~(\ref{eq:r-m}) are used to show that the
total mass loss rate for clouds of mass $m$ in a volume $V_A$ is given
by
\begin{align}
\Big(&\frac{\dot{M}_c}{{\rm M}_{\odot}{\rm Myr}^{-1}}\Big) = -0.44\Big(\frac{M_c}{M_{\odot}}\Big)^{\alpha_c}\Big(\frac{r_{ref}}{{\rm pc}}\Big)\Big(\frac{M_{ref}}{{\rm M}_{\odot}}\Big)^{-\alpha_c}
\nonumber \\ 
&\Bigg(Q\Big(\frac{T_b}{10^6{\rm K}}\Big)^{5/2}+
(1-Q)\Big(\frac{T_a}{10^6{\rm K}}\Big)^{5/2}\Bigg)\,.
\label{eq:lambda}
\end{align}
Under the assumption that $T_b$, the mean temperature of supernova
remnants, and $T_a$, the mean ambient temperature, are constant over
any single timestep we can write
\begin{equation}
\Big(\frac{\dot{M}_c}{{\rm M}_{\odot}{\rm Myr}^{-1}}\Big)\equiv \lambda \Big(\frac{M_c}{M_{\odot}}\Big)^{\alpha_c},
\end{equation}
In order to calculate the constant of proportionality, $\lambda$, we
use an estimate of the mean temperature and density inside of a
supernova remnant.  These estimates were obtained by noting that by
definition $Q\equiv V_B/V_A$. ($V_B$ and $V_A$ represent the total
volume in bubbles and the ambient phase respectively).  The mean
radius of a supernova remnant is then
\begin{equation}
r_b=\Big(\frac{3QV_{A}}{4\pi N_{SN}}\Big)^{1/3}\,,
\end{equation}
where $N_{SN}$ is the total number of supernova explosions that have
affected the local ambient medium (Calculated from equations
\ref{eq:stellarlifetime} and \ref{eq:imf}).  The mean density inside
the supernova remnants, $n_b$, may then be calculated from
Eq~(\ref{eq:nh}) and Eq~(\ref{eq:x}) and the mean temperature from
Eq~(\ref{eq:Tf}).

Over a period of time $\Delta t$ a cloud with mass $M_{I}$ will
evaporate to a mass of $M_F$, given by:
\begin{equation}
M_F = \Big(M_{I}^{(1-\alpha_c)} - \lambda \Delta t\Big)^{1/(1-\alpha_c)}
\end{equation}

\begin{figure}
\begin{center}
\includegraphics[width=8.3cm,clip]{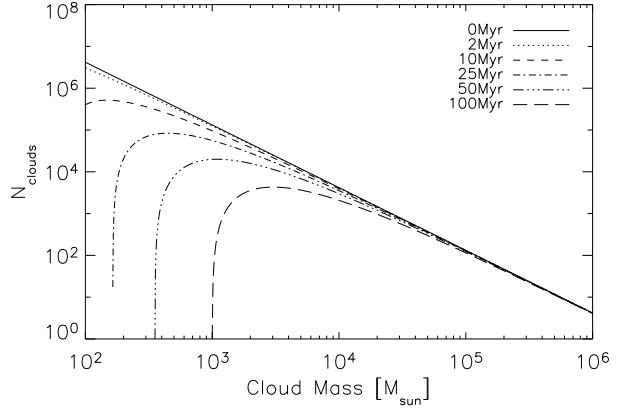}
\end{center}
\caption{Evolution of a population of molecular clouds as they are
evaporated by a hot ambient medium. The initial cloud mass function is
a power law.  The temperature of the ambient medium is assumed to be
$10^5$K, the porosity of the medium is assumed to remain
constant at 0.2, and the temperature of the supernova remnants is
$\approx 10^6$K.  Thermal conduction acts to preferentially destroy
the smaller clouds.}
\label{fig:conductionplot}
\end{figure}

Thermal conduction efficiently destroys smaller clouds, but its
effects are far less dramatic on larger clouds.  Fig
~(\ref{fig:conductionplot}) shows the evolution of an initially power
law mass spectrum of clouds in a hot medium.  The energy used to
evaporate a mass $M_F-M_I$ of cold clouds is removed from the
supernova remnants.

\subsection{Mass Resolution Limits}

The sticky particle model allows particles of all types to change
their mass via processes including merging, thermal conduction and
star formation.  For this reason it is necessary for us to introduce
numerical minimum and maximum masses on all particle types.  We define
at the initial time a characteristic mass resolution for our
simulation, $M_{{\rm char}}$, typically this is set equal to the mass
of the ambient gas particles in the initial conditions. Where more
than one mass of ambient particles is present (for example in the
model galaxies discussed in section \ref{sec:results}) we use the mass
of the gas particles that will be forming most stars.  We then define
minimum and maximum particle masses relative to this characteristic
mass scale.

Ambient gas particles may have their mass decreased by the formation
of molecular clouds.  If the total mass of a gas particle becomes less
than $0.1M_{{\rm char}}$ then it is converted entirely into a cloud
particle.  The ambient gas particles may also have their mass
increased by the process of thermal conduction.  If a gas particle
becomes more massive than $4M_{{\rm char}}$ then it is not allowed to
grow any more, and the evaporated cloud mass is given to a different
particle.  In practice this limit is rarely, if ever, reached as
evaporating cold clouds effectively cools the ambient gas particles so
they become inefficient at thermal conduction.

Sticky particles may decrease their mass by star formation and
evaporation.  If the mass of a sticky particle drops below $0.1M_{{\rm
char}}$ then it is either completely evaporated or completely
converted into stars.  Coagulation may drive the mass of a sticky
particle to be very large.  In practice this is not a real concern
since when a sticky particle becomes very massive the rate at which
its internal clouds coagulate also increases, causing it to form stars
very rapidly.

Stars have a maximum and minimum mass of $4M_{{\rm char}}$ and
$0.1M_{{\rm char}}$.  If a star forms with a mass greater than the
maximum allowed mass it is split into a number of smaller star
particles.  A sticky particle may not form a star with a mass lower
than the minimum allowed mass. In this eventuality then the mass of
the \lq unresolved\rq~ stars is tracked internally by the sticky
particle and added into the next star formation event until the total
mass of stars formed reaches the resolution limit of the simulation.

These particle mass limits keep all particle masses in the range
$0.1M_{{\rm char}}$ to $4M_{{\rm char}}$, which both minimises two
body effects between very massive and very small particles and also
prevents the formation of very many low mass particles, which are
computationally very expensive to evolve.

\section{Parameter Estimation}
\label{sec:parameter-est}
The various physical processes in the star formation and feedback
models each have associated with them physical parameters.  Before we
discuss the properties of our model in detail we discuss how its free
parameters can be constrained.

The free parameters that control the thermal instability and formation
of the molecular clouds are $\rho_{{\rm th}}$ and $T_{{\rm th}}$, the
physical density and temperature at which thermal instability is
allowed to set in and radiative cooling creates molecular clouds.
\cite{wolf95} found that a diffuse ISM naturally settles into two
stable phases, with a sharp cutoff between the ambient and molecular
phases at a density of approximately 1 atom per cm$^3$. We use this as
the value of $\rho_{{\rm th}}$. A threshold temperature $T_{{\rm
th}}=10^5$K allows the gas in galaxies which cools radiatively to
$\sim 10^4 K$ to collapse into clouds but prevents supernova heated
material (typically at temperatures of $10^6 K$) from forming
molecular clouds until it has radiated away most of its supernova
energy.

The properties of the molecular clouds themselves are contained in
four parameters: $r_{\rm ref}$, $M_{\rm ref}$, and $\alpha_c$ as
defined in Eq ~(\ref{eq:r-m}) and $u_c$, the internal energy per unit
mass of molecular clouds.  The first three values are set by
comparison with observations of molecular clouds in the nearby galaxy
M33 (\cite{wils90}):
\begin{equation}
\label{eq:r-m-obs}
\Big(\frac{r_{{\rm c}}}{{\rm pc}}\Big)=(36 \pm 6)\Big(\frac{M}{10^{5} M_{\odot}}\Big)^{0.3\pm 0.1}
\end{equation} 
Thus $r_{\rm ref}$ and $M_{\rm ref}$ are assumed to be 36pc and
$10^5M_{\odot}$ respectively.  This calibration (and an assumed
$\alpha_c$ of 0.3) suggest a radius of $122\pm 6$pc for the largest
clouds observed in the MW ($~6\times 10^6M_{\odot}$ (\cite{will97}).

The properties of the stars and associated feedback are contained
within four parameters: $x$, the slope of the IMF; $E_{51}$, the
energy of each supernova blast in units of $10^{51}$erg; $M_{{\rm
\star,min}}$, the minimum star mass; and $M_{{\rm \star,max}}$, the
mass of the largest allowed stars.  For $E_{51}$ we use the fiducial
value of 1.0 noting, however, that the value of $E_{51}$ is very
uncertain and may be significantly higher.  The effects of varying
$E_{51}$ are investigated in section \ref{sec:away}.  For the purposes
of this work uncertainties in the IMF are neglected and $x$ is assumed
to take on the standard Salpeter value of 1.35. We follow \cite{kawa03}
in adopting values $0.2 M_{\odot}$ and $60 M_{\odot}$ for the minimum
and maximum stellar masses, respectively.

The star formation efficiency in a single cloud collapse is also
somewhat uncertain and is known to be approximately
$\epsilon_\star\approx$ 11\% (\cite{will97}) in the MW.

The thermal conduction efficiency is characterised by two numbers:
$\alpha_{{\rm con}}$, the ratio of the blast wave velocity to the
ambient sound speed and $\phi$, the efficiency of thermal conduction.
Following MO77, the value of $\alpha_{{\rm con}}$ is set to 2.5 (for
the ideal Sedov blast wave case, $\alpha_{{\rm con}}$ is 1.68, the
presence of thermal conduction changes this value).  The thermal
conduction efficiency parameter is assumed to be $\phi=1$.  The
presence of magnetic fields and turbulence may change $\phi$
significantly; we investigate the effect of moving away from this
value in Sect.~\ref{sec:away}

This leaves $v_{{\rm stick}}$ (the maximum relative cloud velocity for
mergers) and $\eta$ (the fraction of a cloud's velocity lost per
non-merger collision) as free parameters that are hard to constrain
via observation.  It is noted that the large scale behaviour of a
given simulation is largely independent of the value of $\eta$.  This
is because the cold cloud velocity dispersion is always limited by
$v_{{\rm stick}}$.  In the following section simple simulations are
used in order to calibrate the properties of the physical model.

\subsection{One Zone Simulations}
\label{sec:onezone}
\subsubsection{Simulation Details}
\label{sec:ozics}

\begin{figure}
\begin{center}
\includegraphics[width=8.3cm,clip]{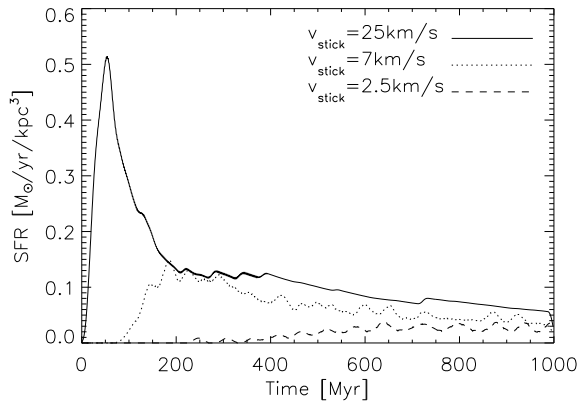}
\end{center}
\caption{Star formation rate as a function of time for a one zone box
with three different values of $v_{{\rm stick}}$.  The initial gas
density is $n_0=2$ cm$^{-3}$ in each case.  Each curve follows the
same general shape, there is an initial delay during which the first
GMCs are forming.  The unopposed collapse of the first GMCs causes a
burst of star formation, which is quickly regulated by the effects of
feedback from stellar winds and supernova explosions.  After this
initial burst the star formation rate in the simulation settles down
and gradually decreases as the gas in the box is used up. }
\label{fig:vm}
\end{figure}

A \lq one zone model\rq\, is a periodic box that represents a fixed
mass and volume (i.e. a static, periodic box with no mass outflow).
The ambient ISM phase is assumed to be homogeneous. Initially, for a
chosen mean density of matter we assume that 50\% of the material is
initially in the hot phase at a temperature of $T_0=10^6$K.  The
remaining gas is initially in cold clouds with an initial mass
function that is a very steep ($N(M)dM \propto M^{-8}dM$) power law.
Numerically we represent the different phases as follows: the ambient
phase is assumed to be homogeneous and isotropic and so is represented
by a single density and temperature throughout the whole periodic
volume, molecular clouds are represented by discrete sticky particles
that are spawned at a random point in the computational volume with a
random velocity, stars are not tracked individually, and are assumed
to heat the whole volume evenly when they undergo SN explosions.  The
mass resolution of the molecular phase is approximately
$10^7M_{\odot}$, although this figure is varied in section
\ref{sec:calib}

This initial situation represents hot, dense gas that has just begun
to experience a thermal instability and started forming its first
molecular clouds.  The volume we simulate is one cubic kpc. The hot
phase will evaporate cold clouds through thermal conduction, and can
cool via radiative processes using a simple tabulated cooling function
from \cite{suth93} (assuming solar metallicity).  Cold cloud particles
are scattered randomly throughout the volume and given random
velocities.  Clouds do not feel gravitational forces. Depending on the
parameters, the ambient phase will cool radiatively to form more
molecular gas. Clouds will coagulate to form GMCs, which in turn form
stars. The associated SNe evaporate smaller clouds, and may heat the
ambient medium and quench the star formation. This sequence of events
is plotted in Fig~(\ref{fig:vm}), the same general shape is observed
for each value of $v_{\rm stick}$, there is a brief delay as the first
clouds coagulate to form GMCs, these clouds them collapse and form
stars, which undergo supernovae and quench the star formation. On a
longer timescale, the quiescent star formation rate slowly decreases
as the available gas is consumed by stars. Since the dynamical
equilibrium is reached on a very short timescale, typically a couple
of hundred Myr, we assume instantaneous recycling when considering
supernova feedback.  The role of $T_{th}$ is suppressed in the one
zone simulations, due to the face that energy injected via supernovae
cannot escape the volume.

The lack of self gravity does not affect the global properties of the
volume significantly. From Eq ~(\ref{eq:sasl}) assuming typical cloud
properties ($M_c \approx 10^5 M_\odot$, $r\approx 50{\rm pc}$) and a
reasonable velocity dispersion ($\sigma_{1D} = 7{\rm km/s}$) the ratio
of the geometric part of the cross section to the gravitationally
focused part is approximately 0.1, therefore direct collisions between
clouds account for the majority of the collisions and gravitational
focusing makes for an effect of only 10\%.  In the following section
we will choose a value of $v_{\rm stick}$ by comparing the star
formation rates in one zone volumes with the Schmidt law, and also
look at the properties of one zone volumes.

As noted in section \ref{sec:parameter-est} the properties of the
simulation are largely independent of $\eta$.  We assume a value of
0.5 throughout the rest of this paper.

\subsubsection{Calibrating the base model}
\label{sec:calib}

\begin{figure}
\begin{center}
\includegraphics[width=8.3cm,clip]{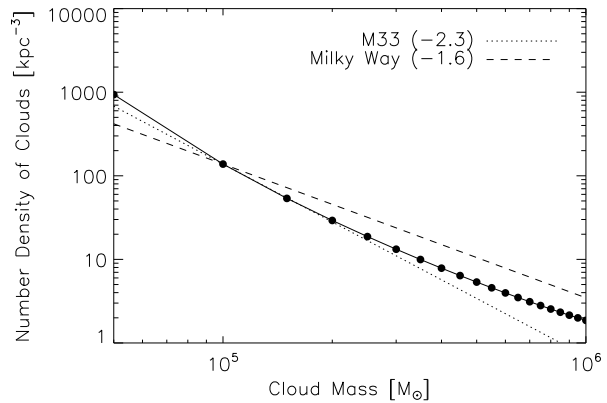}
\end{center}
\caption{Mass function of clouds after 1Gyr in a one zone model.  The
dashed and dotted lines represent the slopes of the mass functions in
the MW (Solomon et al (1987))and M33 (Rosolowsky \& Blitz (2004)).
The numbers in the legend represent the power law slopes in each of
the galaxies.  It is clear that we obtain a good agreement between our
model and the cloud mass spectrum in real galaxies.}
\label{fig:mf}
\end{figure}

\begin{figure}
\begin{center}
\includegraphics[width=8.3cm,clip]{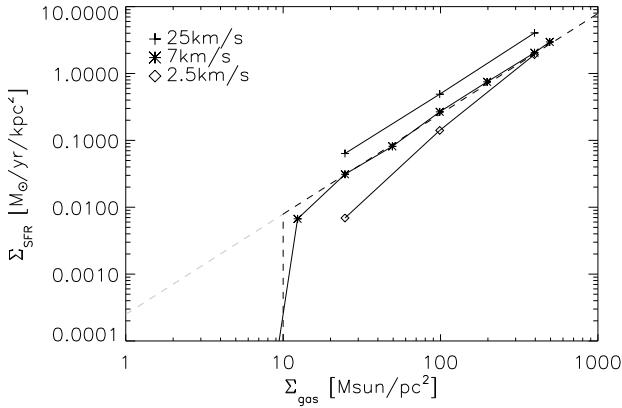}
\end{center}
\caption{Schmidt law.  The diagonal dashed line represents the
observed star formation law (Kennicutt 1998) and the vertical line
represents the observed cutoff in star formation ($10M_{\odot}{\rm
pc}^{-2}$; Schaye 2004).  Each point represents the star formation
rate averaged over a period of 500Myr for a separate one zone
simulation.  Data is shown for three different values of $v_{{\rm
stick}}$, the base value used in all subsequent simulations is
7km/s. We calculate star formation rates by averaging the star
formation rate in the simulation volume over a 500Myr period.  Surface
densities were calculated from volume densities by assuming a disk of
thickness of 1kpc.}
\label{fig:schmidtlaw}
\end{figure}

\begin{figure}
\begin{center}
\includegraphics[width=8.3cm,clip]{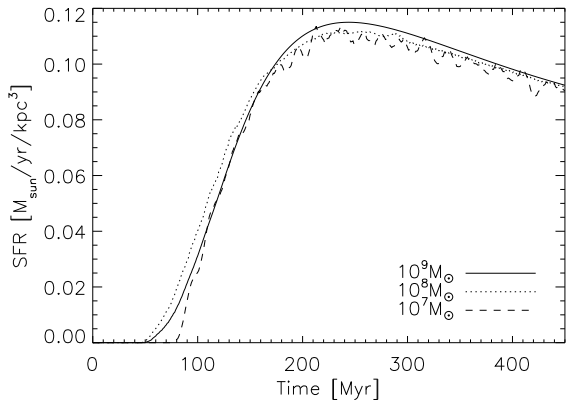}
\end{center}
\caption{Star formation rate as a function of time for one zone models
with three different mass resolutions.  The star formation rate remains
almost unchanged over two orders of magnitude in mass resolution.  The
coarsest mass resolution of $10^9M_\odot$ corresponds to the entire
one-zone system being represented with a single particle with all
clouds interactions modelled with the coagulation equations.}
\label{fig:resolution}
\end{figure}

\begin{figure}
\begin{center}
\includegraphics[width=8.3cm,clip]{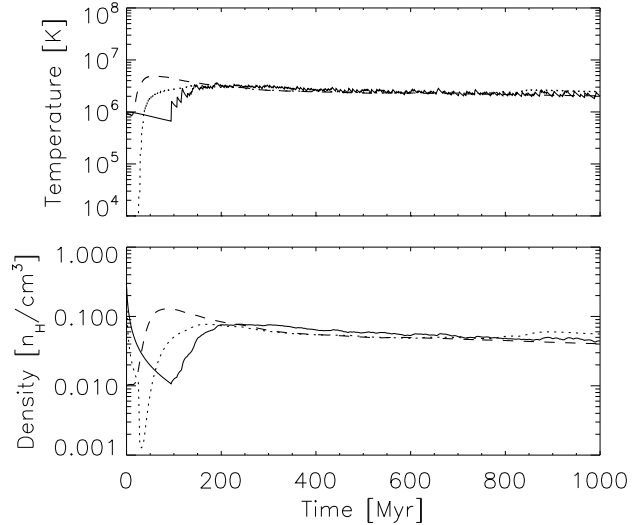}
\end{center}
\caption{Temperature and density of the ambient phase of a one zone
model for a variety of different choices of initial temperature and
density.  The total gas density, ambient gas plus clouds, is always 2
cm$^{-3}$. The interplay of supernova feedback and radiative cooling
quickly brings the system into an equilibrium independent of the
initial value.  }
\label{fig:ozics}
\end{figure}

\begin{figure}
\includegraphics[width=8.3cm,clip]{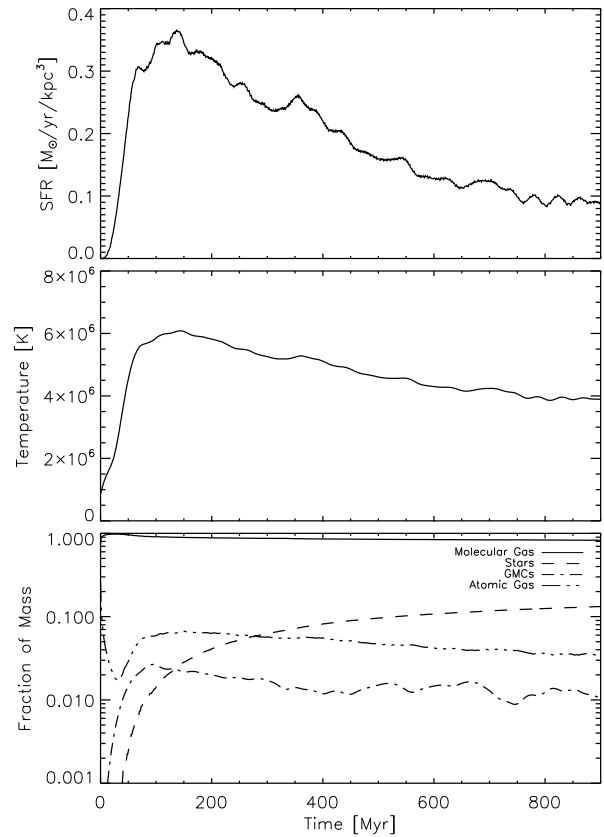}
\caption{The large scale properties of a one zone model with initial density $n_{0}=2$ cm$^{-3}$.  The physical parameters used in this model are the same as the base model as discussed in section \ref{sec:calib}}
\label{fig:phases1}
\end{figure}

\begin{figure}
\includegraphics[width=8.3cm,clip]{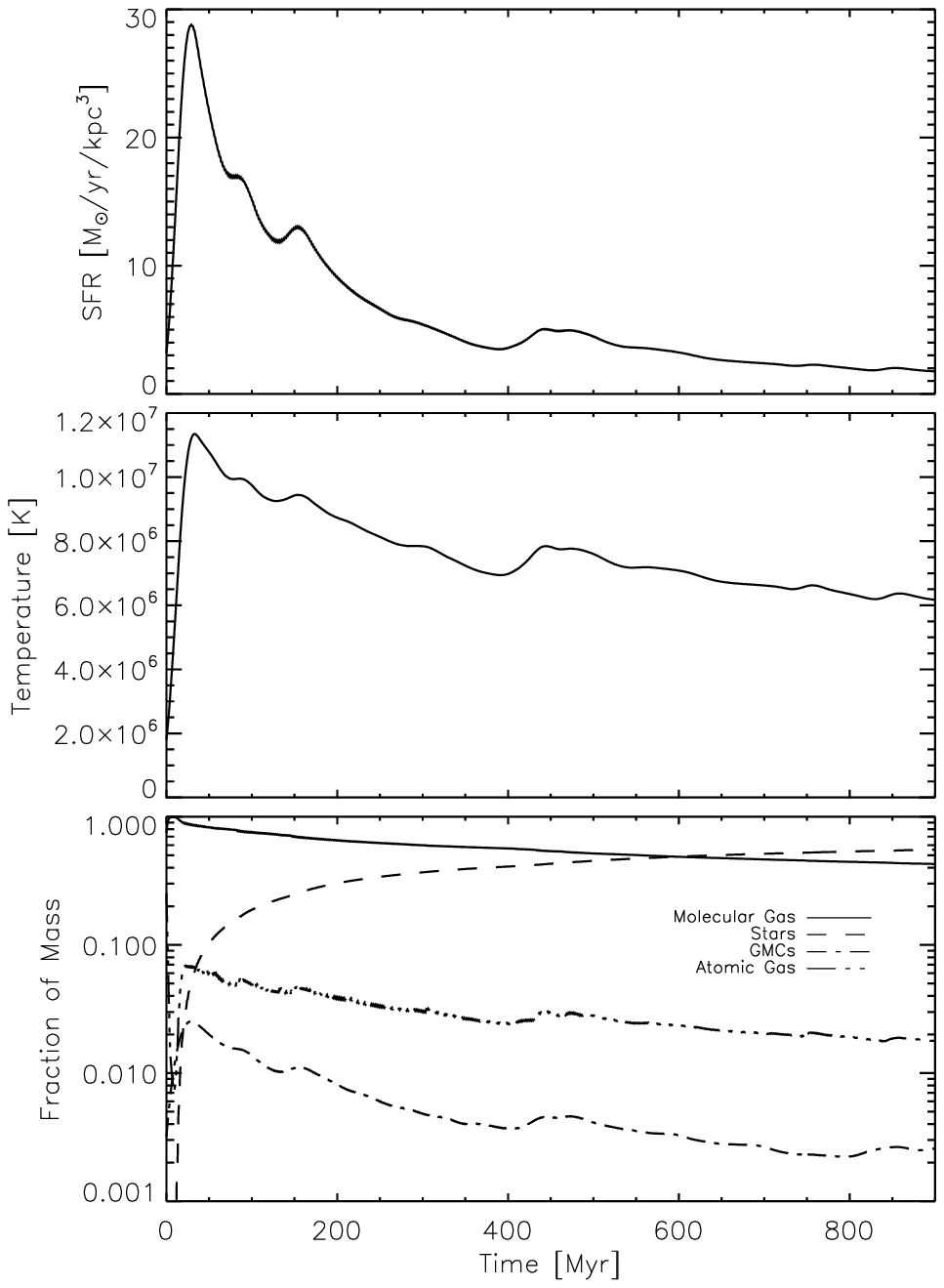}
\caption{Same as Fig ~(\ref{fig:phases1}) except with an initial gas
density of $n_0=16$ cm$^{-3}$}
\label{fig:phases16}
\end{figure}

The one zone model provides a useful sandbox in which we can
investigate a wide variety of parameter choices in a relatively
computationally inexpensive environment. In the following section we
discuss our choices for the values of the different parameters.  The
effects of moving away from this \lq base model\rq~ are discussed more
fully in later sections.

The parameters that are available for tuning the output of the model
are as follows: the cold cloud reference size and radius ($r_{{\rm
ref}}$, $M_{{\rm ref}}$); the slope of the cloud mass-radius relation,
$\alpha_c$; the efficiency of star formation in any given cloud
collapse, $\epsilon_\star$; the maximum relative cloud velocity for
merger ($v_{{\rm stick}}$) and the amount of energy ejected per SNII
event ($E_{51}$).

Even though the initially assumed cloud spectrum, $N(M)dM\propto
M^{-8}\,dM$, is very steep and far from equilibrium, SNe feedback and
cloud coagulation rapidly build a mass spectrum $N(M)dM\propto
M^{-\alpha}\,dM$ with $\alpha\approx 2$ (Fig ~(\ref{fig:mf})), close
to what is the observed cloud spectrum in the MW (dashed line) and M33
(dotted lines). This gives us confidence that, although the modelling
of cloud formation is simple, it does produce a realistic cloud
spectrum.

The ISM model can also reproduce the observed Schmidt law.  We find
that in our model the interaction between the coagulation of clouds
and their destruction by stars leads to a SFR-density relation that is
in good agreement with observation (Fig ~(\ref{fig:schmidtlaw})) if
$v_{\rm stick}$ is set to 7km/s.  This represents a reasonable value
for the molecular cloud velocity dispersion, as considering
theoretical models for the origin of random motion on molecular
clouds, we would expect typical velocity dispersions in the range
5-7km/s (\cite{jog88}).

The effect of changing the mass resolution of the simulation over two
orders of magnitude is demonstrated in Fig ~(\ref{fig:resolution}).
Sub-resolution clouds that are simulated only by integrating the
coagulation equations are designed to behave in exactly the same way as
the resolved cloud particles in the simulation, and so we expect the
simulations not to depend strongly on particle number. This is borne
out by the good agreement between simulations carried out with only one
resolved particle (Fig ~(\ref{fig:resolution}), line with mass
resolution of $10^9 M_{\odot}$) where all of the physics is followed by
integrating the sub-grid equations in a single particle and simulations
with a hundred particles that are followed explicitly. 

As stated in previous sections, the behaviour of a one zone model is
virtually independent of its initial temperature and the fraction of
the gas that starts off in the cold phase.  This behaviour is
demonstrated in Fig ~(\ref{fig:ozics}).  A one zone volume with total
initial density of $n_0=2$cm$^{-3}$ was evolved with a variety of
different initial values for the initial temperature and initial
fraction of the mass in the hot phase.  We observe that regardless of
the initial choices for these two quantities the system quickly
settles down to its equilibrium state.  This process occurs through
the opposing actions of thermal conduction and supernova feedback.

Fig ~(\ref{fig:phases1}) and Fig ~(\ref{fig:phases16}) show the
behaviour of the large scale properties of two different one zone
volumes as a function of time.  The only difference in the initial
conditions of the two one zone volumes is their initial density Fig
~(\ref{fig:phases1}) shows the evolution of a one zone volume with a
total density of 2 atoms/cm$^3$; Fig ~(\ref{fig:phases16}) shows
exactly the same plots for a density of 16 atoms/cm$^3$.  The initial
temperature of the hot phase in both simulations is $10^6$K.  In both
cases the star formation rate follows the same general shape.  There
is a small period of time at the beginning of the simulation where
small clouds are coagulating and there is no star formation.  When
GMCs are formed there is a large burst of star formation that is
quickly quenched by feedback SN and thermal conduction in SN bubbles.
The temperature of the diffuse phase is regulated by a combination of
supernova feedback (acting to increase the temperature) and radiative
cooling.  Due to the fact that we do not allow mass to leave the one
zone volume and also assume instantaneous recycling, the temperature
profile very closely matches that of the star formation rate.  It is
noted that in the one zone simulation with the largest density, the
temperature of the ambient phase is held at a higher temperature by
the action of supernovae.  The fraction of the gas in the molecular
phase is lower in the high density simulation due to the increased
amount of evaporation by thermal conduction in the high temperature
ambient phase.  In the following sections the star formation and
feedback prescriptions are tested in a more realistic situation.

\section{Results}
\label{sec:results}
As a more physically interesting test of the star formation model we
have conducted simulations of various isolated model galaxies.  In
this section we introduce the details of the two different types of
simulation performed and present the properties of the stellar disk
and associated ISM in each case.

\subsection{Quiescent Disk}
\label{sec:isolated}
One of the fundamental properties that a star formation prescription
must be able to reproduce is that in MW like conditions, the resulting
behaviour should be similar to that in the MW.  In this section we
discuss the properties of galaxy simulations set up to approximate the
conditions in the MW's quiescent disk.

\subsubsection{Simulation Details}
\label{sec:isoics}
We set up a simplified model of a MW type galaxy using initial
conditions from GalactICS (\cite{kuij95}).  GalactICS generates near
equilibrium distributions of collisionless particles consisting of a
disk, bulge and halo.  These models consist of a spherical bulge
component; an approximately exponential disk, which is rotationally
supported in the x-y plane and supported by random motion in the z
direction; and an approximately spherical halo.

  We add baryonic material to this distribution by converting the disk
and bulge in their entirety into SPH particles at a temperature of
$10^4K$. 1\% of the material in the halo is converted to baryons with
a temperature of $10^6K$.  The addition of baryonic material puts the
system well out of equilibrium so each simulation is run adiabatically
for 50Myr to allow the galaxy to relax closer to its equilibrium state
before the additional physics is allowed to operate.  The total mass
in the disk, bulge and halo are $1\times 10^{10} M_{\odot}$,
$0.43\times 10^{10} M_{\odot}$ and $1.1\times 10^{11} M_{\odot}$
respectively.  The mass resolution of particles in each of three
realisations of this galaxy are summarised in table
\ref{tab:simulationres}.  These masses were chosen such that the
gaseous particles in each of the three components have approximately
the same mass and the dark matter halo particles have masses as close
as possible to the gas particle mass.  The gravitational softenings
for the disk particles (that is: gas, sticky and star) is set to
0.1kpc, 0.05kpc and 0.02kpc in simulations GAL\_LORES, GAL\_BASE and
GAL\_HIRES respectively.  The dark matter particles have softening
lengths ten times larger than the disk particles.

\begin{table}
  \begin{center}
    \begin{tabular}{llll}
      Resolution ($M_{\odot}$)& $Disk$ & $Bulge$ & $Halo$ \\
      \hline
      Low        &   $6.6\times10^6$ &    $5.84\times10^6$  & $5.6\times10^6$ \\
      Base        &   $8.3\times10^5$ &    $7.5\times10^5$  & $7.1\times10^5$  \\
      High         &   $1.0\times10^5$ &    $9.2\times10^4$  & $1.0\times10^5$  \\
      \end{tabular}
  \end{center}
  \caption{Initial particle masses in three different realisations of the
  model GalactICs galaxy that are used throughout this paper.  The
  disk and bulge consist entirely of baryons, the halo additionally
  contains dark matter.  All masses are in units of $M_\odot$.
  Baryons are added to the dark matter halo by converting a random 1\%
  of the halo DM particles into gas.  The dark matter particle mass in
  the halo is therefore the same as the gas particle mass.}
  \label{tab:simulationres}
\end{table}

The GalactICS simulations provide a test of the code in a situation
somewhat similar to a quiescent MW disk.  As discussed in section
\ref{sec:model}, all simulations were performed with the entropy
conserving SPH code GADGET2 (\cite{spri05}), with all of the physics
discussed in section \ref{sec:model} implemented as additional
modules.  Table \ref{tab:simulationnames} contains a brief summary of
the different simulations.  Typical timestep size in one of the base
simulations is $\sim 10^4$yr, although this figure is smaller at early
times when bursts of supernovae heat gas very strongly.

\begin{table*}
  \begin{center}
   \begin{tabular}{lll}
	  \hline
	  Name & Details & $N_{{\rm gas}}$ \\
	  \hline
	  GAL\_BASE & Base GalactICs model & 19330\\
	  GAL\_LORES & Base model with degraded mass resolution & 2450 \\
	  GAL\_HIRES & Base model with improved mass resolution & 255000 \\
	  GAL\_BASE\_LOSN & $E_{51}$ decreased by factor of 5 & 19330\\
	  GAL\_BASE\_HISN & $E_{51}$ increased by factor of 5 & 19330\\
	  GAL\_BASE\_LOCON & Conduction efficiency decreased by factor of 5 & 19330\\
	  GAL\_BASE\_HICON & Conduction efficiency increased by factor of 5 & 19330\\
	  GAL\_BASE\_LOZ & Gas metallicity set to 0.5 Solar & 19330\\
	  GAL\_BASE\_HIZ & Gas metallicity set to 1.5 Solar & 19330\\
	  ROT\_BASE & Spherical rotating collapse & 15000\\
	  ROT\_LORES & Spherical rotating collapse & 4000\\
	  ROT\_HIRES & Spherical rotating collapse & 45000\\
	  \hline \\
   \end{tabular}
  \end{center}
  \caption{Brief table of simulation references and details. $N_{{\rm
  gas}}$ shows the number of gas particles in the disk, bulge and halo
  combined.}
  \label{tab:simulationnames}
\end{table*}

\subsubsection{The Base Simulations}

\begin{figure}
\includegraphics[width=8.3cm,clip]{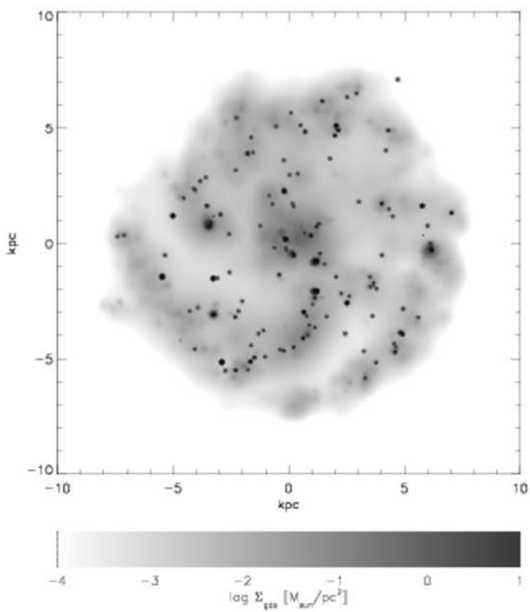}
\caption{Demonstration of the star formation properties of the
isolated galactic disk. The continuous field represents the molecular
gas surface density of the simulated galactic disk, spiral structure
is evident.  The points represent the sites of star formation within
the previous 10Myr.  Most star formation events represent the collapse
of a single GMC, resulting in the formation of $10^{5}M_{\odot}$ of
stars.  It is clear that star formation is occurring primarily in the
spiral arms of the galaxy. }
\label{fig:mapcloud}
\end{figure}

\begin{figure*}
\centerline{\scalebox{0.9}{\includegraphics[width=\textwidth,clip]{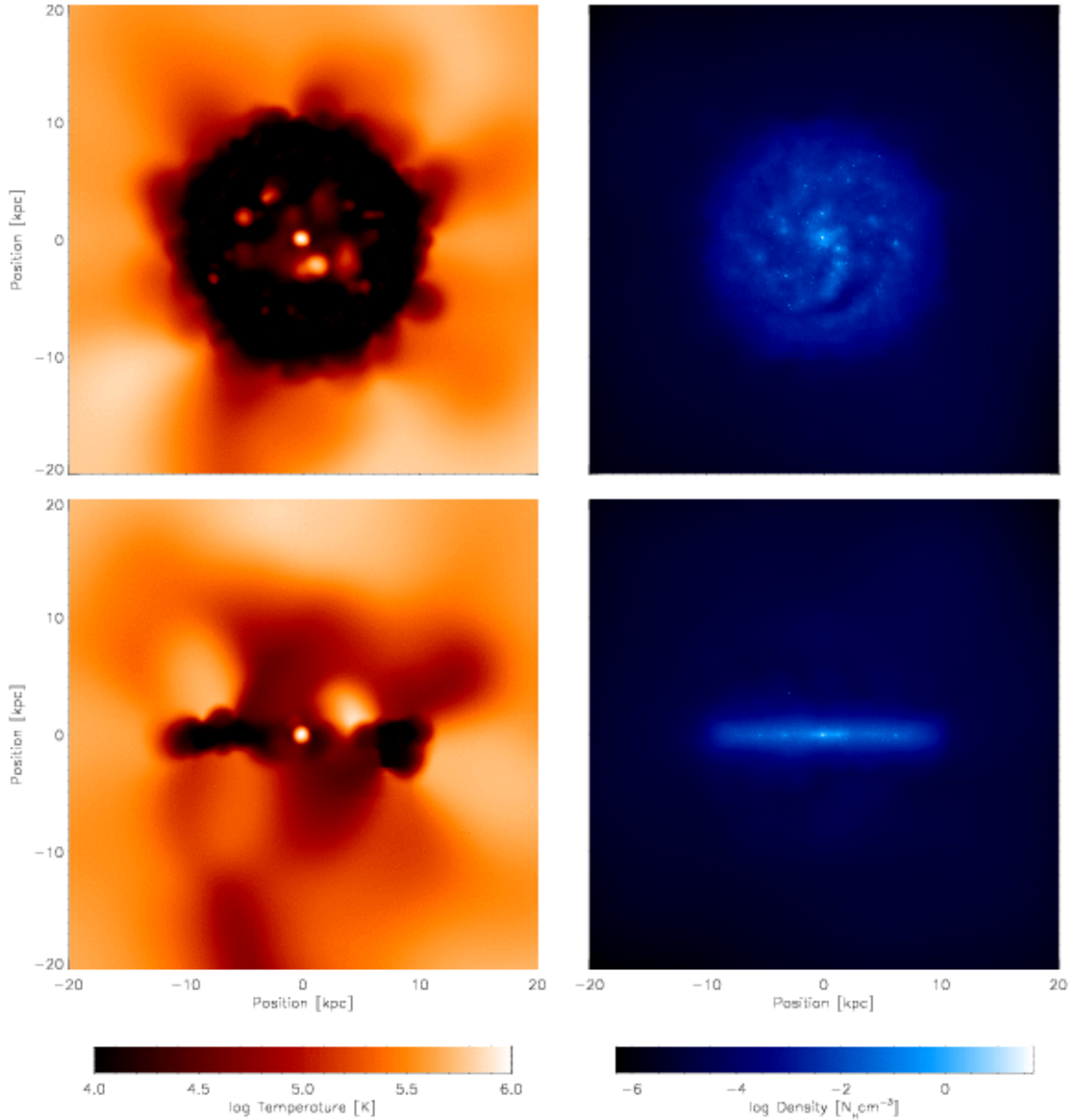}}}
\caption{A thin slice of the gas temperature and density distributions
after 1Gyr in run GAL\_BASE.  The slice is taken directly through the
centre of mass of the stellar disk.  The temperature plot clearly
shows regions of strongly heated gas, these are areas near to sites of
active star formation, where the massive, short lived stars are
heating the ambient material via SN explosions.}

\label{fig:mapedge}
\end{figure*}

\begin{figure}
\begin{center}
\includegraphics[width=8.3cm,clip]{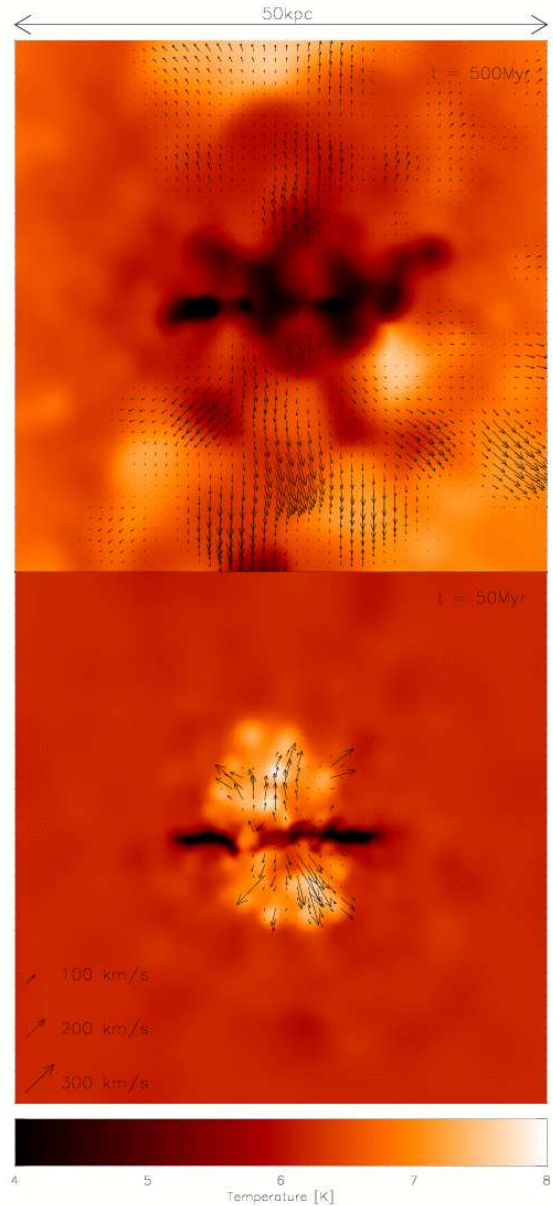}
\end{center}
\caption{A thin slice of the gas temperature field through the centre
of a disk galaxy simulation.  The arrows represent the gas velocity
field, taking into account only gas that has been heated by
supernovae.  The generation of bipolar outflows from the galactic disk
is very clear.  The lower plot represents the galaxy after 50Myr, the
upper panel is the same galactic disk after 500Myr.}
\label{fig:winds}
\end{figure}

\begin{figure}
\includegraphics[width=8.3cm,clip]{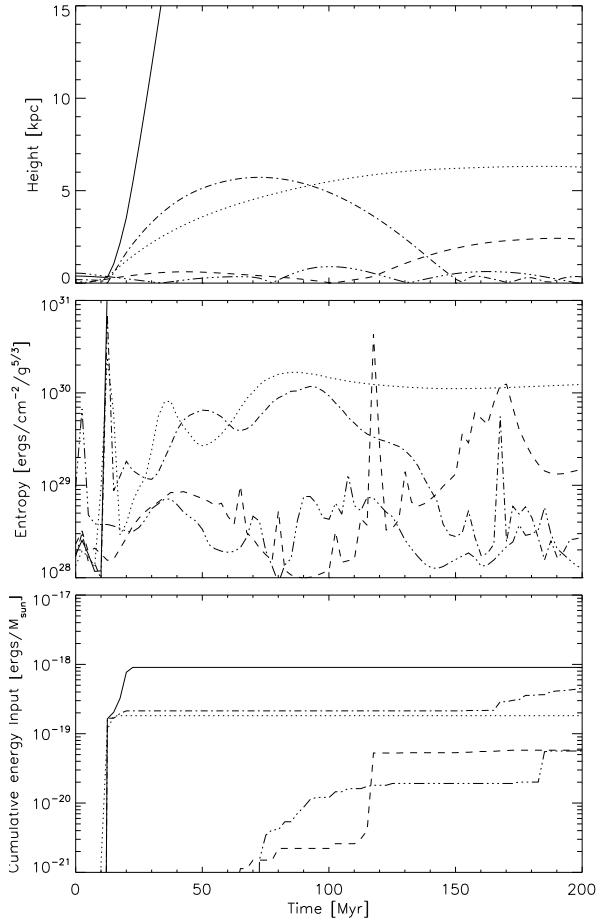}
\caption{Relationship between the number of supernova heatings and the
distance from the mid-plane of the disk for a sampling of five
particles from the simulated galaxy.  Each different linestyle
represents a different SPH particle.  The top panel shows the
perpendicular distance from the centre of the stellar disk, the
central panel shows the entropy of each particle and the lower panel
the cumulative amount of thermal energy that has been dumped into the
particle.  It is clear that some particles with a higher entropy are
lifted away from the galactic disk where they cool and rain back down
on the galactic disk within a hundred Myr of being supernova heated.
Other particles are ejected violently from the galaxy, their density
becomes very low and they evolve adiabatically.}
\label{fig:fountain}
\end{figure}

\begin{figure}
\begin{center}
\includegraphics[width=8.3cm,clip]{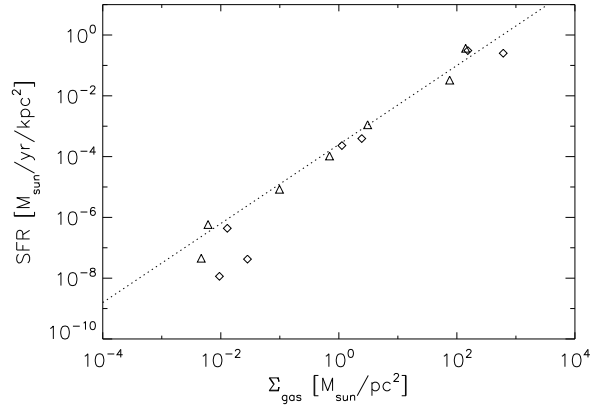}
\end{center}
\caption{The density-star formation rate relationship for our
simulated quiescent disk.  The diamond shaped points represent the
observed star formation rates after 200Myr, and the triangles
represent the star formation rates in the same disk after 1Gyr.  The
dashed line is the observed Schmidt law due to Kennicutt (1998).  The
points on this plot were calculated by taking radial bins, then
calculating the mean density and star formation rate in each bin over
a short period.  The quiescent galaxy follows the observed Schmidt law
throughout its lifetime.}
\label{fig:galacticsschmidt}
\end{figure}

\begin{figure}
\begin{center}
\includegraphics[width=8.3cm,clip]{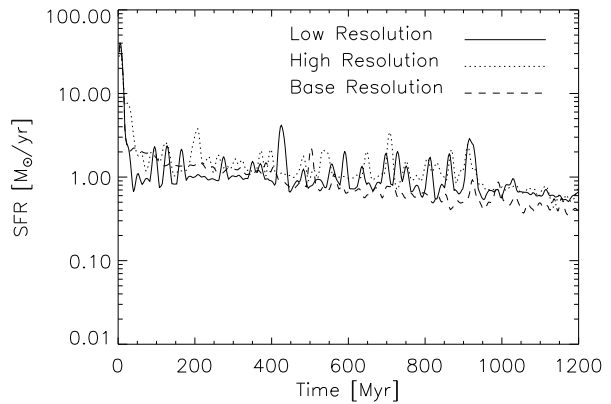}
\end{center}
\caption{Star formation rate as a function of time for isolated galaxy
models with different mass resolutions.  The mass resolution is 64
times better in the high resolution disk than the low resolution one.
The three simulations plotted are GAL\_LORES, GAL\_BASE and
GAL\_HIRES. The fact that the star formation rate remains almost
unchanged shows that numerical convergence has been achieved.}
\label{fig:galresolution}
\end{figure}

\begin{figure}
\begin{center}
\includegraphics[width=8.3cm,clip]{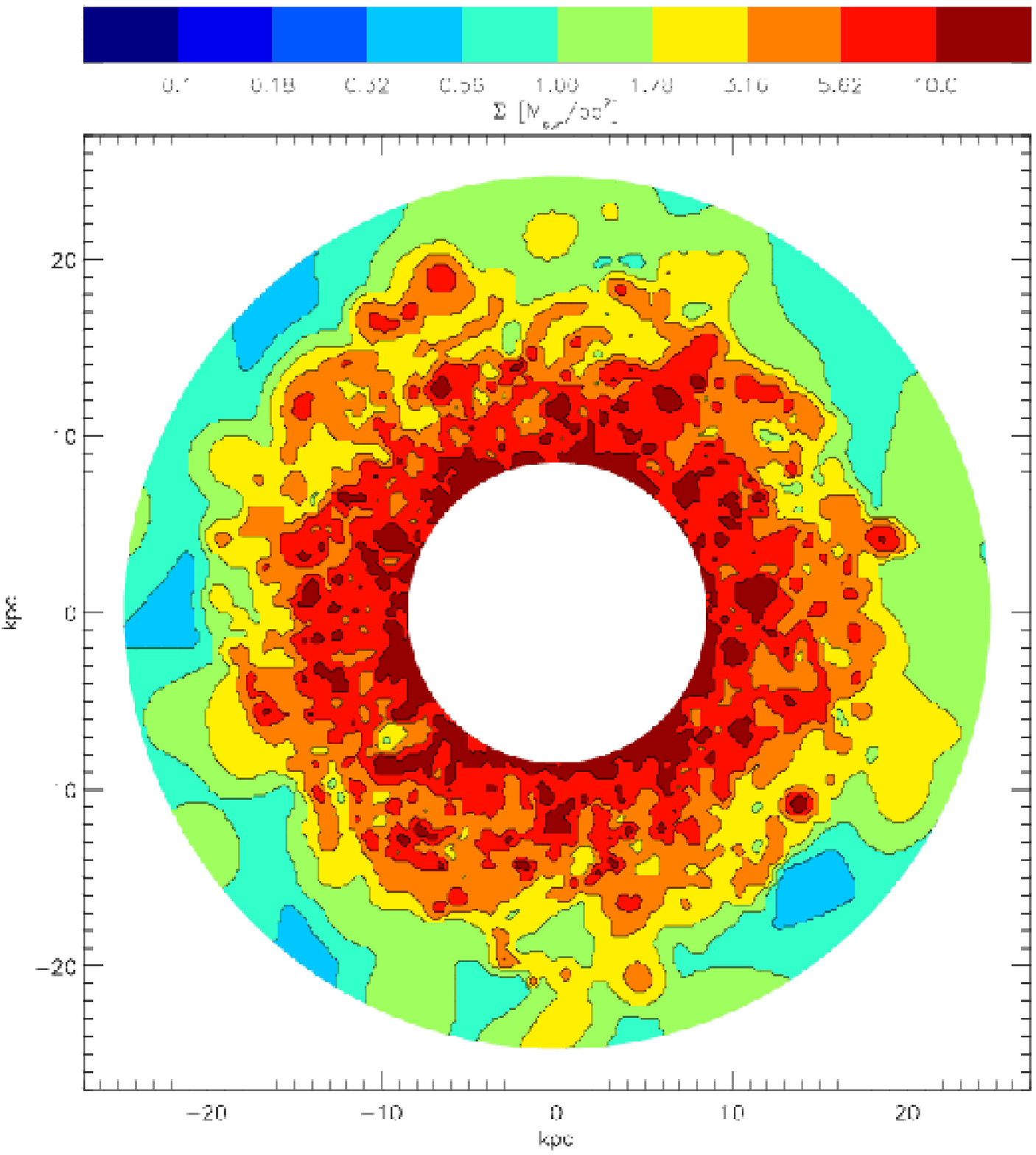}
\end{center}
\caption{Plot of the distribution of gas atomic in a galactic disk
after 500Myr of evolution.  Colours and plot dimensions are matched to
those in Levine et al (2006) for easy comparison to observations.  The
inner circle represents the radius from the centre of the galaxy to
the position of the sun.  The outer radius is where the observations
of Levine et al (2006) are truncated.  The simulated MW has a surface
density profile in close agreement with that observed by Levine et
al.}
\label{fig:gasmap}
\end{figure}

In this section we will discuss simulations run with the base set of
physical parameters (section \ref{sec:calib}).  Most simulations were
run at the base mass resolution as defined in table
\ref{tab:simulationres}.

The large scale behaviour of the model galaxy is as follows:
Immediately after switching on the additional star formation physics
the dense, thermally unstable gas in the disk and bulge collapses into
cold clouds, which quickly cause a large burst of star formation.
After approximately 500Myr the galaxy settles down into a quiescent
state with a star formation rate of approximately 1$M_{\odot}/{\rm
yr}$.  The star formation rate gradually decreases as the cold gas in
the galaxy is consumed by stars.

It is known that in the MW, most areas of active star formation are
concentrated in the galactic spiral arms (e.g. \cite{enga03}).  Fig
~(\ref{fig:mapcloud}) shows that our simulations reproduce this
behaviour.  In the sticky particle model this behaviour occurs
naturally as the converging gas flows in galactic spiral arms lead to
an increased merger rate and, therefore, to the presence of more star
forming clouds.  Face and edge on temperature and density plots of the
standard resolution galaxy are shown in Fig ~(\ref{fig:mapedge}).  The
gas heated by SNII is preferentially situated perpendicular to the
plane of the disk, suggesting that the feedback scheme is
preferentially heating the low density gas and setting up strong
outflows.  Fig ~(\ref{fig:winds}) shows the behaviour of the supernova
heated gas in a thin slice through the centre of the disk.  Initially
there is a strong burst of star formation (lower panel of Fig
~(\ref{fig:winds})), followed shortly by a burst of supernova
explosions that heat the gas around the galactic disk as hot at
$10^8$K.  Most of this gas is driven straight out of the halo in a
direction perpendicular to the galactic disk.  Later on, as the
supernova rate dies down, gas is heated more gently and is ejected
from the galactic disk in the form of a fountain reminiscent of the
galactic fountains present in the MW.  This behaviour is demonstrated
in figure \ref{fig:fountain}, which shows for a random subset of
particles from the gas disk the number of times they have been heated
as a function of time with their height above the galactic disk.  It
is clear that upon being strongly heated, some particles are ejected
from the galactic disk and fall back down a few hundred Myr later.
Others remain in dense regions and cool immediately.  Some escape the
disk completely in the form of a galactic wind.  This behaviour was
also observed in the multiphase star formation models of
\cite{scan06}, suggesting that it is a more general feature of
multiphase models.

Additionally our simulated galaxies are in good agreement with the
observed Schmidt law, Eq ~(\ref{fig:galacticsschmidt}).  This
behaviour arises due to the self-regulation of the simulated ISM.  At
higher densities more molecular clouds are formed and so star
formation rates are higher.  Runaway star formation is prevented by
various forms of stellar feedback, which prevent too many clouds from
forming.  The slope of the Schmidt law in the simulated galaxies may
be changed by altering the value of $v_{{\rm stick}}$.  Higher values
of $v_{{\rm stick}}$ lead to clouds coagulating more rapidly, and so
low density regions of the galactic disk undergo more star formation.
The effect is less severe in the higher density regions of the galaxy
as strong feedback from large bursts of star formation can effectively
regulate the amount of star formation.  Conversely a lower value of
$v_{{\rm stick}}$leads to a steeper Schmidt law with a lower overall
star formation rate.  This behaviour is demonstrated in Fig
~(\ref{fig:vm}) for the one zone model.  The value of $v_{{\rm
stick}}$ used to reproduce the Schmidt law slope and normalisation in
the one zone model also works in the simulated galaxy and the observed
galaxy follows the observed Schmidt law very closely throughout its
whole lifetime (figure ~(\ref{fig:galacticsschmidt})).

The resolution independence of the star formation prescription is once
again demonstrated by Fig ~(\ref{fig:galresolution}).  The star
formation rate between the highest and lowest mass resolution
simulations is in remarkably good agreement.  The general form
followed by all simulations is that there is a strong burst of star
formation at the initial time, this is rapidly quenched by supernova
feedback, and a self-regulating ISM is set up.  The star formation
rate slowly increases as the gas in the galactic disk is either used
up or ejected in the form of winds.

Recent observations of the gas content of the MW have allowed the
construction of maps of its gas surface density (\cite{levi06}).  In
order to compare the properties of our model to observations, another
GalactICs model was generated with properties as close as possible to
those of the MW.  The total mass of the galactic disk was set to
$5\times 10^{10} M_{\odot}$, and the scale radius of the exponential
disk to $4.5 {\rm kpc}$.  This simulation was evolved for 1Gyr. The
resulting gas distribution is shown in Fig ~(\ref{fig:gasmap}).  Our
simulations are in good agreement with the observations of \cite{levi06}.

These properties suggest that the star formation and feedback
prescriptions behave well in a quiescent disk, a more robust test of
how they perform in a more general situation is given by the rotating
collapse simulations.

\subsection{Rotating Collapse}
\label{sec:rotcoll}
\subsubsection{Simulation Details}

The second simulation we investigate is the collapse of a rotating
spherical halo (\cite{nava93}) with an initial $1/r$ density profile
consisting of 90\% collisionless dark matter and 10\% baryonic
material.  The mass of the rotating sphere is
$1\times10^{12}M_{\odot}$ and its initial radius is 100kpc. Velocities
are chosen such that the sphere is initially rotating as a solid body
with a spin parameter of 0.1. Once again this simulation is created at
three different mass resolutions, corresponding to dark matter
particle masses of $5.2\times10^8M_\odot$, $5.8\times10^7M_\odot$ and
$2.0\times10^7M_\odot$, with corresponding gravitational softenings of
1.93kpc, 0.96kpc and 0.68kpc.  The rotating collapse simulations
model, in a crude way, the collapse of a protogalaxy, and allow us to
investigate how the ISM model behaves when it is initially far from
equilibrium and in situations with strong shocks and rapid density
changes.

\subsubsection{The Base Simulations}

\begin{figure}
\includegraphics[width=8.3cm,clip]{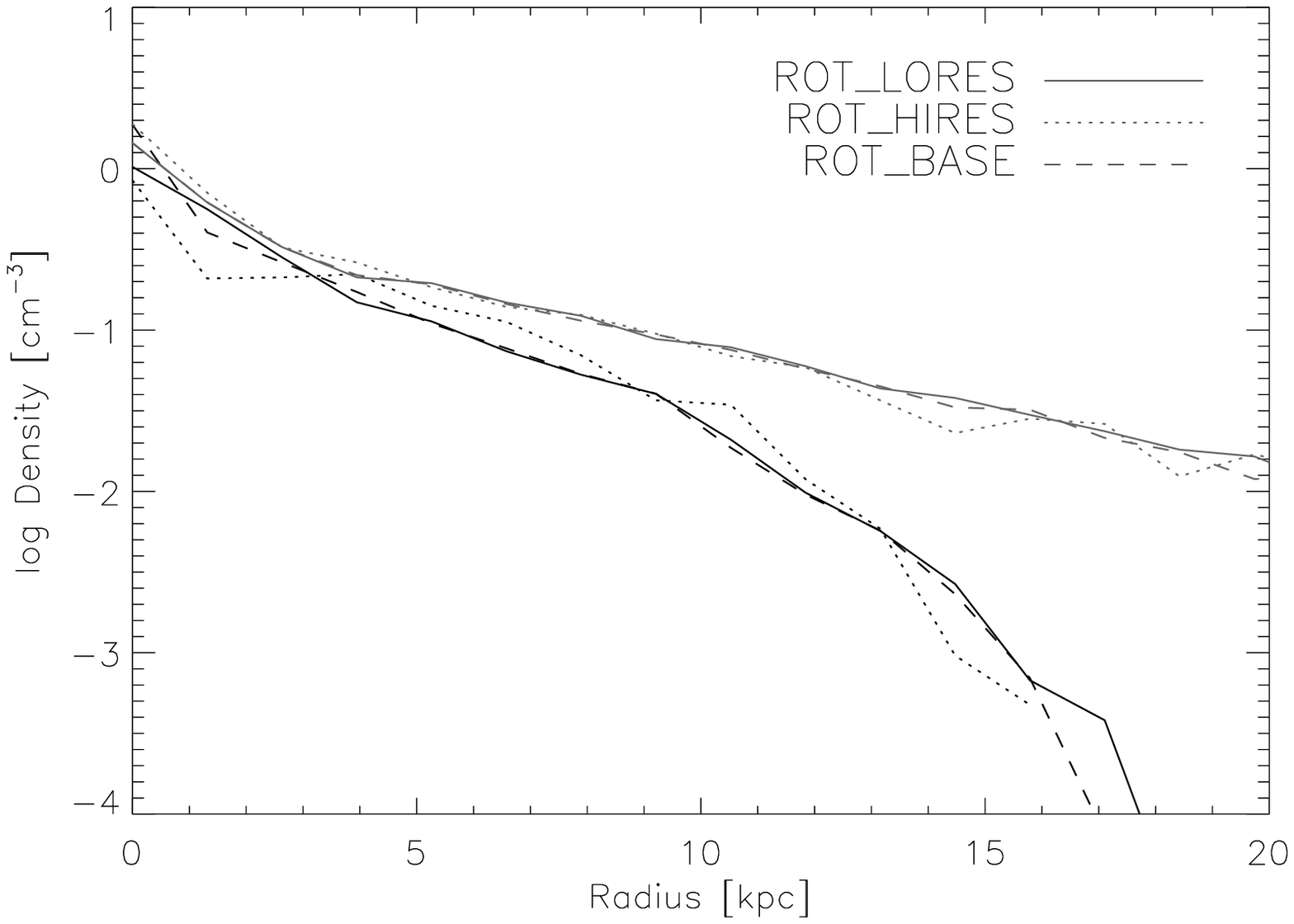}
\caption{Density profiles for the ambient material (grey lines) and
the molecular clouds (black lines). The solid lines represent the low
resolution rotating collapse simulations and the dotted lines
represent the highest resolution simulations.  Agreement between the
high and low resolution simulations is very good.}
\label{fig:densityprofiles}
\end{figure}

\begin{figure}
\begin{center}
\includegraphics[width=8.3cm,clip]{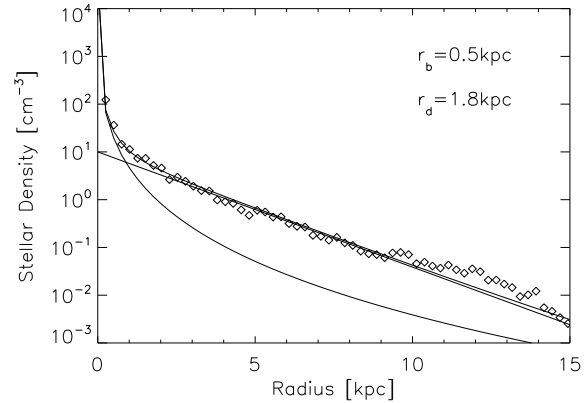}
\end{center}
\caption{Stellar density profile in the base rotating collapse simulation.  Solid lines represent the best fit exponential and a de Vaucouleurs profile.  The scale radii used in the exponential and de Vacouleurs fits are $r_d$ and $r_b$ respectively.  It is clear that even starting from an initial condition far from equilibrium we generate a stellar disk with a surface density profile similar to that in observed galaxies.}
\label{fig:starprof}
\end{figure}

\begin{figure}
\begin{center}
\includegraphics[width=8.3cm,clip]{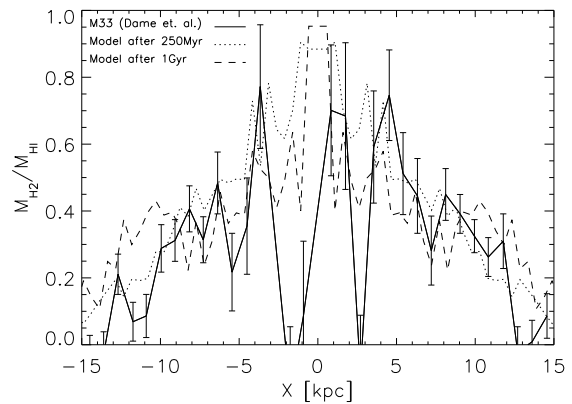}
\end{center}
\caption{Fraction of gas in the molecular phase as a function of
distance from the centre of the galactic disk for the rotating
collapse model at two different times.  The solid line represents the
same data as plotted from M31 (Dame et al (1993)), with the x-axis
representing position along the major axis of the galaxy.  The
simulated galactic disk is in good agreement with observation.}
\label{fig:molfrac}
\end{figure}

\begin{figure*}
\centerline{\scalebox{0.9}{\includegraphics[width=\textwidth]{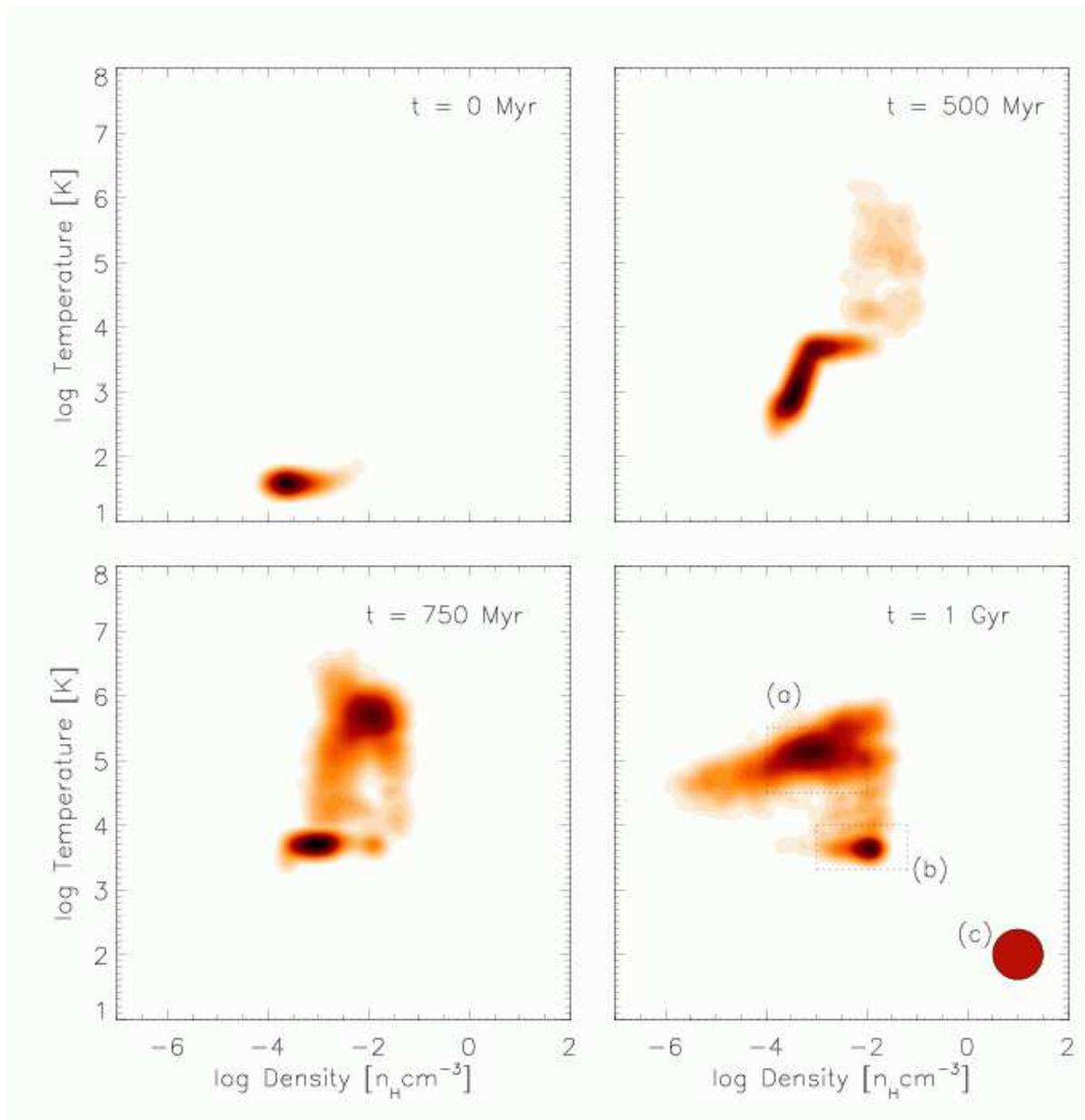}}}
\caption{Density-Temperature relation for a rotating collapse
simulation showing the creation of three distinct components.  In the
final time plot region \emph{a} contains gas in the halo of the
galaxy.  Region \emph{b} is gas in the disk of the galaxy, which
reaches an equilibrium temperature of $\sim10^4K$ and \emph{c}
represents the approximate position of the molecular clouds, at an
assumed temperature of $100K$. The gas in the halo consists both of
gas that failed to cool on to the disk, and gas that was originally in
the disk but has been heated by supernovae.  Approximately 45\% of the
mass is in the two hot phases, 40\% in the disk and 15\% in the cold
molecular clouds.}
\label{fig:rhoT}
\end{figure*}

\begin{figure}
\begin{center}
\includegraphics[width=8.3cm,clip]{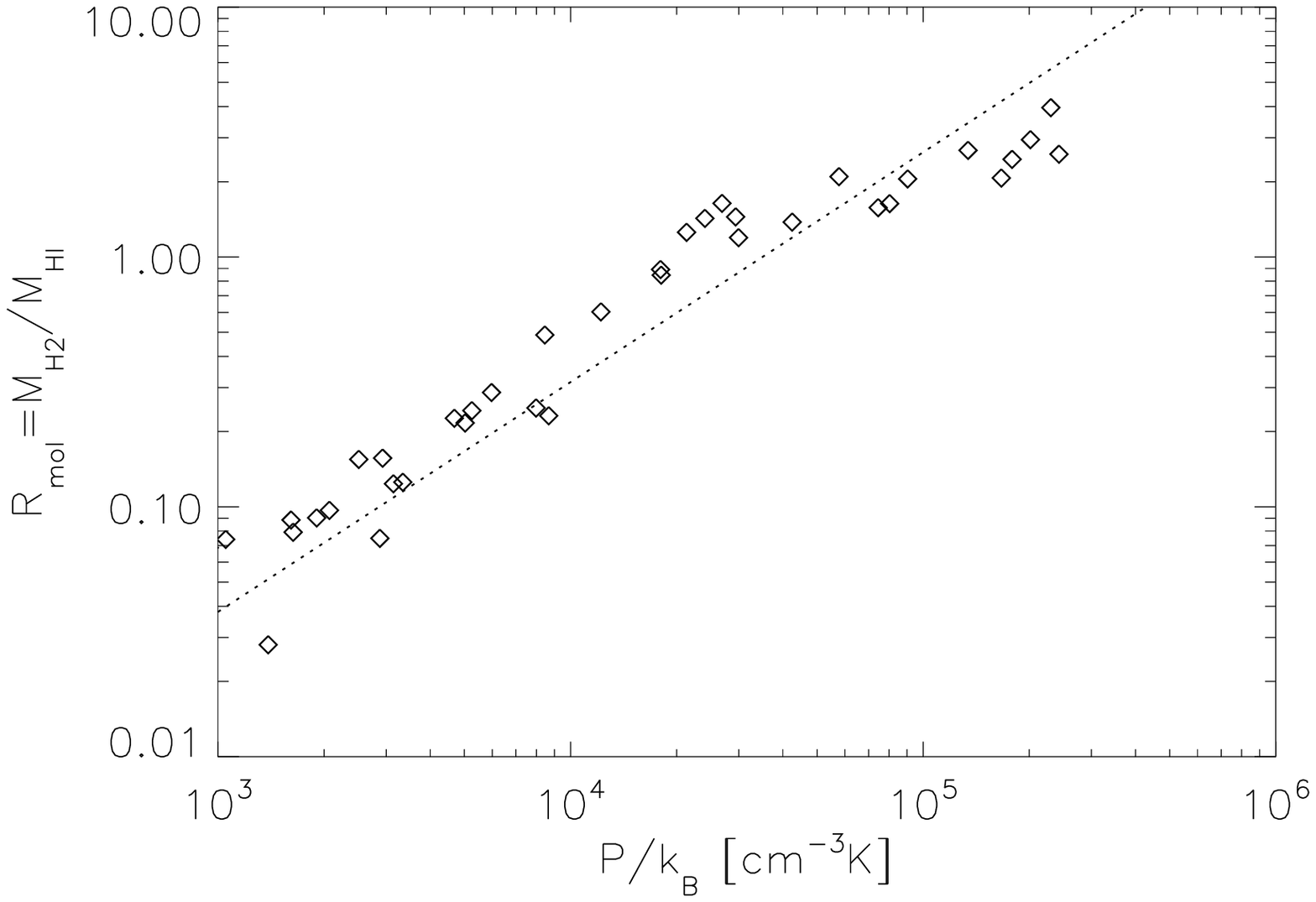}
\end{center}
\caption{The relation between the pressure in the midplane of the
galactic disk and the ratio of gas in the molecular phase to the
atomic phase, $R_{mol}$.  The dotted line represents the observational
result (Eq ~(\ref{eq:rmol}))due to Blitz \& Rosolowsky (2006).  The
simulated galaxy is in good agreement with observation.}
\label{fig:molpres}
\end{figure}

After 2Gyr the density profiles of each of the three phases of matter
are shown in Fig ~(\ref{fig:densityprofiles}) and Fig
~(\ref{fig:starprof}).  The density profiles are averaged around the
disk; each radial bin represents a ring centered on the centre of mass
of the disk.  Figure ~(\ref{fig:densityprofiles}) shows the radial
density profiles of both the ambient and molecular gas.  The three
different resolution simulations once again behave in very similar
ways.  Fig ~(\ref{fig:starprof}) shows the radial density profile of
stellar mass, demonstrating that our star formation prescription gives
rise to an exponential disk and a bulge component well fitted by the
standard $r^{1/4}$ law.

The rotating collapse simulations are especially interesting because
they start far from equilibrium and so features that arise in the
final particle distribution are purely an effect of the physics in the
simulation and not just set up by hand in the initial conditions.  Fig
~(\ref{fig:molfrac}) shows how the molecular fraction (the ratio of
the mass in molecular hydrogen to the mass in atomic hydrogen) varies
as a function of distance from the centre.  Observational data from
M31 (\cite{dame93}) is included as a comparison.

The evolution of the thermal properties of the halo are shown in Fig
~(\ref{fig:rhoT}).  In its initial state all the gas in the halo is
cold.  As the halo collapses it becomes dense and is shock heated.
The gas that ends up in the disk comes to an equilibrium between
radiative cooling and the heating due to SNe at approximately $10^4K$
and a halo of hot, SN heated gas at $\sim 10^6K$ is gradually formed.
The addition of the molecular gas ($T\sim100$K, $\rho\sim 100-1000
{\rm cm}^{-3}$) forms an ISM with four phases: shocked halo gas, SNe
heated material, cold molecular clouds and warm disk gas.  The first
two components are hard to distinguish on the $\rho-T$ plot.

\cite{blit06} have argued that the ratio of molecular to atomic gas in
galaxies, $R_{{\rm mol}}$, is determined by hydrostatic pressure and
through observations of nearby galaxies found the following relation:
\begin{equation}
R_{{\rm mol}}=\Big[\frac{P_{{\rm ext}}/k_B}{(3.5\pm0.6)\times10^4}\Big]^{0.92\pm 0.07},
\label{eq:rmol}
\end{equation}
where $P_{{\rm ext}}$ is an estimate of the mid-plane pressure.
$R_{mol}$ is the ratio of mass in the atomic and molecular phases.
Fig ~(\ref{fig:molpres}) shows the data for the ROT\_BASE simulation
after 1 Gyr alongside the observed best fit line (Eq
~(\ref{eq:rmol})).  We use the SPH estimate of the pressure at
$z=0$. To calculate $R_{\rm mol}$ we use all matter within a vertical
distance of 1kpc from the centre of mass of the disk and bin radially.
The model of the ISM clearly reproduces the observed behaviour.

\subsection{Away From the Base Model}
\label{sec:away}
The determination of some of the physical parameters used in our model
is somewhat uncertain.  In this section we investigate the effect of
varying some of the physics included in the models.  Large
uncertainties are present in the determination of some of the
parameters including $E_{51}$, the thermal conduction efficiency, and
the physics we include in our treatment of SN explosions.  In addition
our simulations do not contain a detailed treatment of metals.  We
demonstrate the effects of changing each of these parameters on the
large scale properties of simulated quiescent disks.

The value of $E_{51}$ may differ greatly from unity, for example due
to radiative cooling of the SN remnant.  We investigate the effect of
moving $E_{51}$ away from unity and also look at changing the physics
included in our analytic model for blast wave evolution, firstly by
extending the simple Sedov solution with the fitting formula due to
\cite{tang05} and secondly by investigating the effects of preventing
radiative cooling in supernova remnants.

Altering the value of $E_{51}$ has, as expected, two main effects.
Firstly an increased supernova efficiency can eject gas from the
galactic disk more efficiently and quenches star formation very
quickly .  Secondly, the gas disk in simulations with higher supernova
efficiency is found to be less centrally concentrated; the reverse is
also true in simulations with a low supernova efficiency. This is a
result of supernova feedback ejecting gas from the galactic disk more
efficiently in the runs with a high value of $E_{51}$.  We find that
the fiducial value of $E_{51}=1.0$ provides a good match with the
observed properties of disk galaxies.

More accurate modelling of the evolution of supernova blast waves (by
using the \cite{tang05} fit to the blast wave evolution) does not
significantly change the properties of the galaxy.  Assuming that a
typical supernova remnant expands for approximately $\sim$0.3 Myr
before being dispersed, that the mean ambient temperature of the ISM
is $10^6$ K and the mean density is $10^{-2}$ atoms per cm$^3$ then
the difference in the blast wave radius between the pure Sedov
solution (\ref{eq:sedov}) and the modified fitting formula due to
\cite{tang05} (Eq ~(\ref{eq:tang})) is never more than 20\%.  This is
demonstrated in Fig \ref{fig:sedov}, which shows a pure Sedov blast
wave compared with the simulated result from an SPH simulation of a
blast wave in a hot ($10^6 K$) medium.

The net result of having larger blast waves is that the porosity of the
ISM increases at a greater rate and the delay between a supernova
explosion occurring and its thermal energy being injected into the
ambient medium is decreased.

Finally we can switch off the radiative cooling in supernova remnants.
The effects here are much more dramatic.  In the dense galactic disk a
supernova remnant can typically radiate away over 90\% of its thermal
energy before being disrupted.  Switching off radiative cooling in
supernova remnants, therefore, has a very severe effect in our galaxy
simulations, and leads to the almost immediate suppression of star
formation and much of the material in the galactic disk is ejected
from the galaxy.  This demonstrates that radiative losses from
supernova remnants are of crucial importance in our models.

Our simulation code does not contain a detailed description of mass
feedback from supernovae. Therefore we need to verify that the
expected evolution in metallicity over the timescale of the simulation
will not substantially affect the properties of the galactic disk.

Following \cite{harf06} we use a simple analytic model to estimate the
change in metallicity over the timescale of a typical simulation then
run simulations with metallicites bracketing this range.  By assuming
that stars form at a constant rate of $1M_{\odot}/{\rm yr}$, that
there is one type 2 supernova event per 125$M_{\odot}$ of stars
formed, and using metal yields due to \cite{woos95} we estimate
that over the 1Gyr timespan of one of the quiescent disk simulations
the average metallicity of the galaxy should not change by more than
$0.04Z_{\odot}$.

The base simulations have a metallicity of $1.0Z_{\odot}$, two
additional simulations were run with metallicities of $0.5Z_{\odot}$
and $1.5Z_{\odot}$, far outside the metallicity evolution range
expected in our quiescent disks.  The total amount of stars after 1Gyr
in the high metallicity run is 5\% higher than in the base run.  The
low metallicity run contains approximately 5\% less stars than the
base run. This trend arises because the radiative cooling rate of the
ambient phase is related to the ambient gas metallicity. All the
properties of the three simulated disks agree to within 10\% with the
properties of the base simulations.  The reason for the surprisingly
small dependence of the galaxy properties on the metallicity of the
gas is that in the quiescent disk the gas is maintained at a
temperature of $\sim10^4$K.  At this temperature the radiative cooling
function does not change very much with metallicity.  We do, however,
note that a full treatment of the metal enrichment of the gas is
necessary in performing fully cosmological simulations since here the
metallicity of the gas may significantly affect the way in which it
collapses (e.g. \cite{scan05}).  In these simulations we have
additionally neglected the change in stellar lifetime with metallicity
(see e.g. \cite{rait96}).  Since we are simulating a quiescent disk,
it is not expected that changes to the lifetimes of massive stars will
have a large effect on the properties of the galactic disk.

However, as noted by \cite{harf06} a detailed prescription for the
yields from supernovae is necessary if we want to simulate the early
evolution of a galaxy.

One of the most poorly constrained parameters is the efficiency of
evaporation of molecular clouds through thermal conduction.  Magnetic
fields and turbulence may affect the amount of thermal conduction by a
large amount (MO77).  We ran quiescent disk simulations with the
efficiency of thermal conduction moved by a factor of five in each
direction (GAL\_BASE\_LOCON; GAL\_BASE\_HICON).  More efficient
thermal conduction leads to a lower density of molecular clouds in the
galactic disk, as well as making the cloud mass function more shallow.
The star formation rates are affected by a similar amount, in the
simulations with a high thermal conduction rate the star formation
rate is depressed by an order of magnitude.  As discussed in section
\ref{sec:rotcoll} the base value for the thermal conduction efficiency
reproduces many of the observed properties of the MW.

\section{Conclusions}
\label{sec:conclusions}

Motivated by the fact that we cannot reasonably resolve the Jeans
scale for molecular clouds in galaxy simulations we have introduced a
new star formation and feedback prescription.  We model the ambient
phase of the ISM using a hydrodynamic simulation code and the
unresolved molecular gas using a sticky particle prescription.  Our
model leads to a tightly self-regulating multiphase ISM.  The
multiphase nature of our star formation prescription avoids a lot of
the problems of overcooling that were present in the first generation
of star formation models.  With the exception of the parameter that
controls the molecular cloud coagulation timescale, $v_{{\rm stick}}$,
all the parameters in our model can be tightly constrained by
observation, leaving the cloud coagulation timescale as a free
parameter that we can adjust to match the observed properties of
galaxies.  Where possible our model of the ISM has been formulated in
such a way that the results of a simulation should be independent of
mass resolution.  We demonstrated that the large scale properties of
our simulations were unchanged over a two orders of magnitude shift in
mass resolution.

We have applied the sticky particle star formation model to three
different types of simulation: a simple one zone model, the rotating
collapse of a gas and dark matter sphere and a model of a quiescent
galactic disk.  After using the one zone simulation to set the value
of the parameters that cannot be determined observationally, the sticky
particle model can be applied to the other simulations without any
parameter changes.

The simulations of a quiescent disk galaxy reproduce the observed
Schmidt law with a slope of 1.4 due to the opposing effects of cloud
coagulation and feedback effects. The galaxy also developed a natural
three component ISM. Finally we observe supernova heated gas in the
galaxy being ejected from the disk either in the form of a galactic
fountain, or, when the star formation rate (and associated supernova
rate) is sufficient, in the form of strong bipolar outflows.  Both of
these results arise as a natural consequence of the physics included
in our star formation prescription.

Simulations of the collapse of a rotating sphere of dark matter and
gas reproduced, many of the observed properties of galactic disks,
beginning from an initial condition well out of equilibrium.  In
particular we observe a stellar disk with distinct bulge and disk
components, well fitted by the standard exponential and de Vacouleurs
density profiles.  The fraction of molecular gas in the disk as a
function of radius is reproduced, and agrees well with recent
observations of nearby galaxies.  The observed relation between the
disk midplane pressure and the fraction of molecular clouds is also
reproduced.  We also observe star formation rates comparable to those
in disk galaxies and note that our model reproduces the formation of
stars in the spiral arms of the galaxy.

Our preferred values for most of the parameters are discussed in
section \ref{sec:parameter-est}; and are usually constrained by observation.
We varied the values of those parameters that are only weakly
constrained and foind that our preferred value of $E_{51}=1$ find that
we obtain reasonable results.  Including different physical
descriptions of the evolution of supernova blast waves makes only a
modest difference to our results.  Finally we note that in the
quiescent disk simulation our results depend relatively weakly on
metallicity; although this will not be the case in fully cosmological
simulations, for which a full treatment of metal enrichment via
supernova feedback is necessary.

A natural continuation of this work is to extend our investigations to
higher redshift through the use of fully cosmological simulations, and
to explore the behaviour of the ISM in colliding galaxies.  This work
is currently being pursued.

\section*{Acknowledgements}
CB and TT thank PPARC for the award of a research studentship, and an
Advanced Fellowship, respectively.  TO acknowledges financial support
from the Japan Society for the Promotion of Science for Young
Scientists (1089). We are grateful to Volker Springel for providing us
with the GADGET2 code, to Joop Schaye, Claudio Dalla Vecchia, and Rob
Wiersma for providing us with tabulated radiative cooling and heating
rates and to Peter Thomas, Pierluigi Monaco, Gustavo Yepes and Richard
Bower for useful discussions.  Finally we thank the anonymous referee
for their careful reading of the manuscript, which has substantially
improved the logical flow and clarity of the paper.  All simulations
were performed on the Cosmology Machine at the Institute for
Computational Cosmology in the University of Durham.

\appendix

\section{List of Symbols}
\label{app:list-of-symbols}

$\alpha_{\rm c}$    : Slope of the molecular cloud mass-radius relation. Eq \ref{eq:r-m} \\
$c_{\rm h}$         : Sound speed of the ambient gas phase \\
$\epsilon_\star$    : Fraction of a GMC converted to stars in a collapse. \\
$\eta$              : Fraction of cloud velocity lost to 'cooling' collision \\
$E_{51}$            : Energy ejected per SnII in units of $10^{51}$ergs \\
$E_{{\rm b}}$       : Total energy in a supernova blast wave \\
$E_{{\rm m}}$       : Total kinetic energy in molecular clouds of mass $m$ in a given volume \\
$f_{\rm cl}$         : Filling factor of cold clouds \\
$f_{\rm m}(\sigma_{1},\sigma_{2})$ :  Fraction of collisions between clouds with velocity dispersions $\sigma_{1}$ and $\sigma_{2}$ that lead to mergers \\
$K(m,m')$           : The kernel for aggregation of clouds of masses $m$ and $m'$. Eq. \ref{eq:k} \\
$\lambda$           : Constant of proportionality relating cloud mass and destruction rate by thermal conduction. Eq. \ref{eq:lambda} \\
$\Lambda_{\rm N}$   : Normalised radiative cooling rate \\
$\Lambda_{\rm net}$ : Net radiative cooling rate (ergs cm$^{-3}$s$^{-1}$)\\
$M_{\rm c}$         : Mass of a molecular cloud \\
$M_{\rm char}$      : Characteristic mass resolution of a simulation \\
$M_{\rm ref}$       : Reference cold cloud mass. Eq \ref{eq:r-m} \\
$M_{{\rm \star,min}}$ : Minimum allowed star mass \\
$M_{{\rm \star,max}}$ : Maximum allowed star mass \\
$n_{\rm b}$         : Density internal to a supernova remnant in atoms / ${\rm cm}^3$
$n_{\rm c}$         : Density of a molecular cloud in atoms / ${\rm cm}^3$ \\
$n_{\rm h}$         : Density of the ambient medium in atoms / ${\rm cm}^3$ \\
$N_{\rm SF}$        : The slope of the schmidt law. Eq \ref{eq:schmidtlaw}\\
$n(m,t)$            : The number of clouds with masses between $m$ and $m+dm$     \\
$N(m,t)$            : The number density of clouds with masses between $m$ and $m+dm$     \\
$\phi$              : Efficiency of destruction of cold clouds by thermal conduction \\
Q                   : Porosity of the interstellar medium. Sec. \ref{sec:snnumerics} \\
$r_{\rm c}$         : Radius of a molecular cloud \\
$r_{\rm ref}$       : Reference cold cloud radius. Eq \ref{eq:r-m} \\
$r_{\rm b}$         : The radius of a spherical blast wave \\
$\rho_{\rm c}$      : Mean density of molecular clouds contained in a volume \\
$\rho_{\rm h}$      : Mean density of ambient gas contained in a volume \\
$\rho_{\rm th}$     : Density at which ambient gas becomes thermally unstable \\
$\rho_{\rm SFR}$    : Volume density of star formation \\
$T_{\rm b}$         : Mean temperature inside of a supernova remnant \\
$T_{\rm c}$         : Internal temperature of cold clouds \\
$T_{\rm h}$         : Temperature of the ambient gas phase \\
$u_{\rm b}$         : Thermal energy per unit mass of supernova remnants \\
$u_{\rm c}$         : Thermal energy per unit mass of the cold clouds \\
$u_{\rm h}$         : Thermal energy per unit mass of the ambient phase \\
$\Sigma$            : Cross section for collision between clouds. Eq. \ref{eq:sasl}\\
$\Sigma_{\rm cond}$ : Efficiency of thermal conduction. Eq \ref{eq:sigmacond}\\
$v_{\rm app}$   : Relative approach velocity of two molecular clouds \\
$v_{{\rm stick}}$   : Maximum relative velocity for cloud merger \\
$x$             : Slope of the stellar IMF \\

\section{Energy Transfer Through Coagulation}
\label{app:co-derive}
  Our formalism to treat the evolution of the mass function of clouds
internal to each of the 'multiple cloud' particles will start from the
Smoluchowski equation of kinetic aggregation (\cite{smol16})
\begin{eqnarray}
\label{eqn-smoluchowski}
\frac{\partial n}{\partial t}=\frac{1}{2V}\int^{\infty}_{0}n(m' ,
t)n(m-m',t)K(m',m-m') dm' \nonumber \\
 - \frac{1}{V}n(m,t)\int^{\infty}_{0}n(m',t)K(m,m')dm',
\end{eqnarray}
where $n(m,t)$ represents the number of clouds with masses between $m$
 and $m+dm$ contained within a volume, $V$ and $K(m,m')$ represents
 the kernel for aggregation of clouds with masses $m$ and $m'$, as defined by Eq ~(\ref{eq:k}).

  The fraction of collisions between clouds of masses $m_{1}$ and
$m_{2}$ that lead to mergers is given by
\begin{align}
\label{eqn-merge-frac}
f_{m}(\sigma_{1},\sigma_{2})=&\frac{1}{\sigma_{1}\sqrt{2\pi}}
   \int^{\infty}_{-\infty}
   e^{-\Big(\frac{v_{1}}{\sqrt{2}\sigma_{1}}\Big)^{2}}
\nonumber \\ 
   &\Big[{\rm erf}\big(\frac{v_{1}+v_{{\rm stick}}}{\sqrt{2}\sigma_{2}}\big)
   -{\rm erf}\big(\frac{v_{1}-v_{{\rm stick}}}{\sqrt{2}\sigma_{2}}\big)\Big]dv_{1}\,.
\end{align}

Using this definition of $f_{{\rm m}}$ the Smoluchowski equation becomes

\begin{align}
\label{eqn-smol}
\frac{\partial n}{\partial t}=&\frac{1}{2V}\int^{\infty}_{0}n(m' ,
t)n(m-m',t)K(m',m-m')
   \nonumber \\
& f_{m}(\sigma_{m'},\sigma_{m-m'}) dm' 
   \nonumber \\
- &\frac{n(m,t)}{V}\int^{\infty}_{0}n(m',t)K(m,m')f_{m}(\sigma_{m},\sigma_{m'})dm'\, .
\end{align}

As discussed in section \ref{sec:coagulation}, clouds of mass $m$ may
gain or lose kinetic energy in three ways: clouds of mass $m'$ and
$m-m'$ may merge to form extra clouds of mass $m$.  Clouds of mass $m$
may merge with clouds of any other mass to decrease the number of
clouds of mass $m$.  Finally clouds of mass $m$ can interact
gravitationally with any other clouds, thus losing kinetic energy.
These three processes are termed gain, loss and cooling.

Gain processes may be represented in the following way, where we have
integrated over $m'$ such that the two particles that merge have
masses that sum to $m$
\begin{align}
\dot{E}_{{\rm gain}}& =  
\int^{\infty}_{0}\int^{v_{1}=\infty}_{v_{1}=-\infty}\int^{v_{2}=v_{1}+v_{{\rm stick}}}_{v_{2}=v_{1}-v_{{\rm stick}}}\Big[P(v_{1})P(v_{2})n(m',t)
\nonumber \\
&n(m-m',t)K(m',m-m')f_{m}(m',m-m')E_{f}\Big]
\nonumber \\
&dv_{2}dv_{1}dm'\,.
\label{eqn-deltae}
\end{align}
$P(v_1)$ and $P(v_1)$ are the probability distributions velocities
$v_1$ and $v_2$ and are assumed to be gaussian with standard deviation
$\sigma_1$ and $\sigma_2$ respectively. , $E_{{\rm f}}$ represents the
final kinetic energy of a collision between particles of masses $m'$
and $m-m'$. $E_{{\rm f}}$ is evaluated by considering conservation of
momentum,
\begin{equation}
E_{{\rm f}} = \frac{1}{2}\frac{(m'v_{1} + (m-m')v_{2})^{2}}{m^{2}}\,.
\end{equation}
Eq ~(\ref{eqn-deltae}) then becomes
\begin{align}
\label{eqn-deltae2}
\dot{E}_{{\rm gain}}& = \frac{n(m',t)}{2\pi} 
\int^{\infty}_{0}n(m-m',t)K(m',m-m')
\nonumber \\
&\int^{\infty}_{-\infty}\int^{v_{1}+v_{{\rm stick}}}_{v_{1}-v_{{\rm stick}}}
\frac{1}{\sigma_{m'}\sigma_{m-m'}}
e^{-\Big(\frac{v_{1}}{\sqrt{2}\sigma_{m'}}\Big)^{2}}
e^{-\Big(\frac{v_{2}}{\sqrt{2}\sigma_{m-m'}}\Big)^{2}}
\nonumber \\
&\frac{1}{2}\Big(\frac{(m'v_{1} + (m-m')v_{2})^{2}}{m}\Big)dv_{2}dv_{1}dm'\,,
\end{align}
and Eq ~(\ref{eqn-deltae2}) may be written
\begin{align}
\label{eqn-deltae_gain}
\dot{E}_{{\rm gain}} &= \frac{n(m',t) }{2\pi}
\int^{\infty}_{0}n(m-m',t)K(m',m-m')
\nonumber \\
&\int^{\infty}_{-\infty}\frac{1}{\sigma_{m'}\sigma_{m-m'}}
e^{-\Big(\frac{v_{1}}{\sqrt{2}\sigma_{m'}}\Big)^{2}}
\nonumber \\
&\int^{v_{1}+v_{{\rm stick}}}_{v_{1}-v_{{\rm stick}}}
e^{-\Big(\frac{v_{2}}{\sqrt{2}\sigma_{m-m'}}\Big)^{2}}
\frac{1}{2}\Big(\frac{(m'v_{1} + (m-m')v_{2})^{2}}{m}\Big)
\nonumber \\
&dv_{2}dv_{1}dm'\,
\end{align}

The total kinetic energy of particles of mass $m$ may also be decreased
by mergers between particles of mass $m$ and any other mass (the second
process in the list).  Similarly to Eq ~(\ref{eqn-deltae2}), the
rate of energy loss may be written
\begin{align}
\dot{E}_{{\rm loss}} & = \frac{n(m,t) }{2\pi}
\nonumber \\
&\int^{\infty}_{0}n(m',t)K(m,m')\int^{\infty}_{-\infty}\frac{1}{\sigma_{m}\sigma_{m'}}e^{-\Big(\frac{v_{1}}{\sqrt{2}\sigma_{m}}\Big)^2}
\nonumber \\
&\int^{v_{1}+v_{{\rm stick}}}_{v_{1}-v_{{\rm stick}}}e^{-\Big(\frac{v_{2}}{\sqrt{2}\sigma_{m'}}\Big)^2}\frac{mv_{1}^{2}}{2}
dv_{2}dv_{1}dm'\,.
\label{eqn-deltae_loss}
\end{align}
Finally, the total energy of particles with mass $m$ may be decreased
by collisions between particles of mass $m$ and particles of any other
mass that occur at relative velocities greater than $v_{{\rm stick}}$.  In
this case, the velocity of both particles is decreased by a factor
$\eta$ relative to the centre of mass.  For a collision between
particles of masses $m_{1}$ and $m_{2}$ (velocities $v_{1}$ and
$v_{2}$) the final velocity of particle 1 (denoted $v'_{1}$) is
evaluated by conservation of momentum
\begin{align}
v'_{1} = &\eta(v_{1}-v_{{\rm com}}) + v_{{\rm com}}
\nonumber \\
v_{{\rm com}} = &\frac{m_{1}v_{1} + m_{2}v_{2}}{m_{1}+m_{2}}
\nonumber \\
\Delta E = &\frac{1}{2}m_{1}v_{1}'^{2} - \frac{1}{2}m_{1}v_{1}^{2}\,,
\end{align}
Using these definitions, the change in energy of a particle of mass
$m_{1}$ by gravitational cooling with a particle of mass $m_{2}$,
denoted $\epsilon$ is given by
\begin{align}
\epsilon = \frac{m_{1}}{2}\Big[v_{1}^2(1-\alpha^2)-v_2^2\beta^2-v_{1}v_{2}\alpha\beta\Big],
\end{align}
where
\begin{align}
&\alpha =\eta+\frac{m_{1}}{m_{1}+m_{2}}(1-\eta)
\\
&\beta = \frac{m_{2}}{m_{1}+m_{2}}(1-\eta)\,.
\end{align}
In a similar way to Eq ~(\ref{eqn-deltae2}) the energy loss via
this process may be written
\begin{align} 
\label{eqn-deltae_cool}
\dot{E}_{{\rm cool}} &= \frac{n(m,t) }{2\pi}
\int^{\infty}_{0}n(m',t)K(m,m')
\nonumber \\
&\int^{\infty}_{-\infty}\frac{1}{\sigma_{m}\sigma_{m'}}
e^{-\Big(\frac{v_{1}}{\sqrt{2}\sigma_{m}}\Big)^{2}}
\nonumber \\
&\int^{|v_{1}-v_{2}|>v_{{\rm stick}}}
\epsilon e^{-\Big(\frac{v_{2}}{\sqrt{2}\sigma_{m'}}\Big)^2}
dv_{2}dv_{1}dm'\,.
\end{align}

\section{The Solution of the Coagulation Equations}
\label{app:co-solve}

In our simulations we solve the discrete versions of Eq
~(\ref{eqn-smol}), Eq ~(\ref{eqn-ke}) and Eq ~(\ref{eqn-sigma}), where
by assuming that cloud mass is quantised into $N$ bins characterised
by an index, $i$, where $M_{i}=iM_{0}$ we can write

\begin{equation}
\dot{n}_{k}=\frac{1}{2V}\sum_{i+j=k}K_{ij}f^{m}_{ij}n_{i}n_{j}-\frac{n_{k}}{V}\sum_{j=1}^{j_{max}}K_{jk}f^{m}_{ij}n_{j},
\end{equation}

\begin{equation}
\dot{E}_{k}= \dot{E}_{{\rm gain}} - \dot{E}_{{\rm loss}} - \dot{E}_{{\rm cool}},
\end{equation}

\begin{equation}
\dot{\sigma}_{k}=\frac{\dot{E}_{k}-\frac{1}{2}\sigma_{k}^{2}M_{k}\dot{n}_{k}}{\sigma_{k}n_{k}M_{k}},
\end{equation}

where subscripts represent different mass bins.  $K_{ij}\equiv
K(M_{i},M_{j})\equiv K(iM_{0},kM_{0})$.  The superscript $m$
represents that $f$ is a cross section for particle mergers

To demonstrate the technique for solving these equations we will
consider the numerical solution of the simple Smoluchowski (Eq
~(\ref{eqn-smoluchowski})), which when written in a discrete form
takes on the following form
\begin{equation}
\label{eqn-disc_smol}
\dot{n}_{i}=\frac{1}{2V}\sum^{i-1}_{j=1}n_{j}n_{i-j}K_{ij}-
\frac{n_{i}}{V}\sum^{N}_{j=1}n_{j}K_{ij}.
\end{equation}
Following \cite{bens05} Eq ~(\ref{eqn-disc_smol}) can be rewritten in
the form of a matrix equation
\begin{equation}
\label{eqn-disc_matrix}
\mathbf{\dot{n}}=\mathbf{B\cdot k},
\end{equation}
the vector $k$ has $N\times N$ elements corresponding to
$K(m_{i},m_{j})$.  The kernel matrix, $\mathbf{B}$ has $N \times N
\times N$ elements and may be written more explicitly as
\begin{equation}
\label{eqn-disc_solv}
\dot{n}_{i}=\sum_{jk}B_{ijk}k_{jk},
\end{equation}
where
\begin{equation}
\label{eqn-disc_b}
B_{ijk}=\frac{n_{j}n_{k}}{V}\Big(\frac{1}{2}\delta_{i,j+k}-\delta_{ik}\Big)
\end{equation}
$\delta$ represents a Kronecker delta function.  We solve Eq
~(\ref{eqn-disc_matrix}) implicitly using an iterative method.

The solution of the equations that govern energy exchange between
clouds (Eq ~(\ref{eqn-deltae_gain}), Eq ~(\ref{eqn-deltae_loss}) and
Eq ~(\ref{eqn-deltae_cool})) is the same as for the solution of the
Smoluchowski equation in that we will write the equations in the form
of the linear multiplication of two matrices and then solve this
equation implicitly.  In order to simplify the notation in this
section we will denote the terms in the three equations that are
inside of the integrals over velocity as $\xi$.  Explicitly for the
case of the equation for energy gain (Eq ~(\ref{eqn-deltae_gain})):
\begin{align}
\xi^{G}(m,m')&=\frac{1}{2\pi}
\int^{\infty}_{-\infty}\frac{1}{\sigma_{m}\sigma_{m'}}
e^{-\Big(\frac{v_{1}}{\sqrt{2}\sigma_{m}}\Big)^2}
\nonumber \\
&\int^{v_{1}+v_{{\rm stick}}}_{v_{1}-v_{{\rm stick}}}
\frac{1}{2}
e^{-\Big(\frac{v_{2}}{\sqrt{2}\sigma_{m'}}\Big)^2}
\Big(\frac{(mv_{1} + m'v_{2})^{2}}{m}\Big)
dv_{2}dv_{1}
\end{align} 
The corresponding terms in the equations for energy
loss(\ref{eqn-deltae_loss}) and cooling (\ref{eqn-deltae_cool}) are
denoted $\xi^{L}(m_{1},m_{2})$ and $\xi^{C}(m_{1},m_{2})$
respectively.  Note that the definitions of $\xi$ include the factors
of $2\pi$ and $\frac{1}{\sigma}$ from throughout the equations.

The equation for the total evolution of the energy of a system of
coagulating and cooling particles may be written in terms of these new
functions as:
\begin{align}
&\dot{E}(m)=\int^{\infty}_{0}n(m',t)n(m-m',t)K(m',m-m')\xi^{G}(m',m-m')dm'
\nonumber \\
&-n(m,t)\int^{\infty}_{0}n(m',t)K(m,m')\xi^{L}(m,m')dm'
\nonumber \\
&-n(m,t)\int^{\infty}_{0}n(m',t)K(m,m')\xi^{C}(m,m')dm'
\end{align}
Which when discretized and rearranged becomes 
\begin{equation}
\label{eqn-app-disc}
\dot{E}_{i}=
\sum^{i-1}_{j=1}n_{i}n_{j}K_{ij}\xi^{G}_{ij}
-n_{i}\Big(\sum^{N}_{j=1}n_{j}
K_{ij}\big(\xi^{C}_{ij} + \xi^{L}_{ij}\big)\Big)
\end{equation}
The subscripts represent different mass bins ($n_{j}\equiv n(jM_{0})$).
Our goal is to rewrite Eq ~(\ref{eqn-app-disc}) in the form of a
linear multiplication of two matrices
\begin{equation}
\mathbf{\dot{E}}=\mathbf{C}\cdot\mathbf{k},
\end{equation}
where $\mathbf{k}$ is defined in the same way in the solution of the
Smoluchowski equation, that is: $k_{ij}\equiv K(m_{i},m_{j})$.  The
form of $\mathbf{C}_{ijk}$ that is consistent with Eq
~(\ref{eqn-app-disc}) is given by:
\begin{equation}
C_{ijk}=n_{j}n_{k}\Big(\delta_{i,j+k}\xi^{G}_{jk}-
\delta_{ik}\big(\xi^{C}_{jk} + \xi^{L}_{jk}\big)\Big)
\end{equation}
This form for $\mathbf{C}_{ijk}$ is functionally equivalent to
$\mathbf{B}_{ijk}$ (Eq ~(\ref{eqn-disc_b})) so the solution may
proceed in exactly the same way as for the Smoluchowski equation, the
only difference is the form of the matrix $\mathbf{B}$

The calculation of the quantities $\xi^{G}$, $\xi^{C}$ and $\xi^{L}$
is computationally very expensive so they are initialised once into a
lookup table at the start of every simulation and obtained by bilinear
interpolation thereafter.

\bsp

\label{lastpage}

\end{document}